\documentclass[%
 preprint,
 amsmath,amssymb,
prfluids,
]{revtex4-2}
\usepackage{hyperref}
\usepackage{graphicx}
\usepackage{dcolumn}
\usepackage{bm}
\usepackage{epstopdf, epsfig}
\usepackage{natbib}
\usepackage{array, booktabs, makecell}
\usepackage{siunitx}
\usepackage[version=4]{mhchem}
\graphicspath{{./new_fig/}}

\newif\ifcomments 
\commentsfalse
\commentstrue 

\usepackage[dvipsnames]{xcolor}
\usepackage{xspace}
\renewcommand{\l}{\left}
\renewcommand{\r}{\right}

\newcommand{\revision}[1]{{#1}}

\begin{document}

\title{Physics-Informed Machine Learning of the Lagrangian Dynamics of Velocity Gradient Tensor}

\author{Yifeng Tian}
\email{yifengtian@lanl.gov}
\affiliation{
Computational Physics and Methods Group, Computer, Computational and Statistical Sciences Division (CCS-2), Los Alamos National Laboratory, Los Alamos, NM 87545, USA
}%

\author{Daniel Livescu}
\email{livescu@lanl.gov}
\affiliation{
Computational Physics and Methods Group, Computer, Computational and Statistical Sciences Division (CCS-2), Los Alamos National Laboratory, Los Alamos, NM 87545, USA
}%
\author{Michael Chertkov}
\email{chertkov@arizona.edu}
\affiliation{Program in Applied Mathematics, University of Arizona, Tucson, AZ 85721, USA }

\date{\today}

\begin{abstract}
	
\revision{Reduced models describing the Lagrangian dynamics of the Velocity Gradient Tensor (VGT) in Homogeneous Isotropic Turbulence (HIT) are developed under the Physics-Informed Machine Learning (PIML) framework. We consider VGT at both Kolmogorov scale and coarse-grained scale within the inertial range of HIT. 
Building reduced models requires resolving the pressure Hessian and sub-filter contributions, which is accomplished by constructing them using the integrity bases and invariants of VGT. The developed models can be expressed using the extended Tensor Basis Neural Network (TBNN) introduced in \citep{ling2016reynolds}. Physical constraints, such as Galilean invariance, rotational invariance, and incompressibility condition, are thus embedded in the models explicitly. Our PIML models are trained on the Lagrangian data from a high-Reynolds number Direct Numerical Simulation (DNS). To validate the results, we perform a comprehensive out-of-sample test. We observe that the PIML model provides an improved representation for the  magnitude and orientation of the small-scale pressure Hessian contributions. Statistics of the flow, as indicated by the joint PDF of second and third invariants of the VGT, show good agreement with the "ground-truth" DNS data. A number of other important features describing the structure of HIT are reproduced by the model successfully. We have also identified challenges in modeling inertial range dynamics, which indicates that a richer modeling strategy is required. This helps us identify important directions for future research, in particular towards including inertial range geometry into TBNN.}

\end{abstract}

\maketitle


\section{\label{sec:intro}Introduction}

\revision{Properties of the Velocity Gradient Tensor (VGT), at both small-scale and coarse-grained at a scale from the inertial interval, determine many important characteristics of turbulence \citep{meneveau2011lagrangian}. Bare VGT represents the Kolmogorov (viscous) scale of turbulence and statistics of the energy dissipation, while VGT coarse-grained at a larger scale provides information about respective velocity increments and energy cascade. The aforementioned statistics are highly intermittent and rich in information about dynamics and transfer of energy and interaction among turbulent eddies 
\citep{eyink2006multi,johnson2017turbulence}. The analysis of VGT allows one to shed light on the mechanism of energy cascade, which can be characterized by the geometrical features and highly nonlinear interactions of vorticity and strain, via the so-called  
vortex stretching and strain self-amplification \citep{doan2018scale,carbone2020vortex,johnson2020energy}.} \citet{perry1987description,chong1990general} have proposed an approach to classify local flow topology and structure using invariants of the VGT. The Lagrangian dynamics of the VGT have been studied following the evolution of the invariants along the Conditional Mean Trajectories (CMT) of the flow. The analysis has been performed using data extracted from Direct Numerical Simulation (DNS) corresponding to various fully developed turbulent flows, such as isotropic turbulence, turbulent boundary layer, mixing layers, and more complex flows, such as shock-turbulence interaction \citep[e.g.][]{chong1998turbulence,martin1998dynamics,ooi1999study,wang2012flow,chu2013topological,bechlars2017evolution,tian2019density,das2020characterization,tom2021exploring,carbone2020symmetry}. To accurately represent Lagrangian dynamics of the VGT, reduced empirical models have been proposed to capture the important dynamics without solving the full Navier-Stokes equations. The empirical model reduction approach is justified by our (so far principal) inability to resolve the non-local nature of the turbulence dynamics in a mathematically accurate way, which is true with regards to both Eulerian and Lagrangian attempts of building reduced models rigorously. \citet{meneveau2011lagrangian} provided a comprehensive review of Lagrangian approaches towards building reduced empirical models of turbulence. Restricted Euler Equation (REE) \citep{vieillefosse1982local,vieillefosse1984internal} is the simplest Lagrangian model, and has formed a basis for further exploration. It was shown in \citet{cantwell1992exact} that the REE allows exact integration, also revealing the model's significant handicap --- an un-physical blow-up (singularity) in finite time, due to misrepresentation of the pressure hessian contribution to the VGT dynamics. A number of models were proposed to resolve the finite-time singularity problem, such as linear diffusion model \citep{martin1998diffusion}, stochastic diffusion model \citep{girimaji1990diffusion}, Lagrangian tetrad model \citep{chertkov1999lagrangian}, Lagrangian Linear Diffusion Model \citep{jeong2003velocity}, multi-scale model \citep{biferale2007multiscale} and Recent Fluid Deformation (RFD) approximation \citep{chevillard2006lagrangian,chevillard2008modeling}. More recently, \citet{wilczek2014pressure} have modeled the pressure Hessian (and viscous contributions which were also omitted in the REE modeling) utilizing a random Gaussian field, and then the RFD model was extended in \citet{johnson2016closure}, leading to the Recent Deformation of Gaussian Fields (RDGF) model. \revision{Models based on expansion about tensor bases have been explored in \citep{lawson2015velocity,leppin2020capturing}, where the scalar coefficients (of the tensor basis expansion) are learned from DNS data and/or derived from physical constraints \citep{betchov1956inequality}. \citet{johnson2017turbulence,pereira2018multifractal} developed models to describe small-scale intermittency in HIT at high-Reynolds numbers.} We emphasize that these models were theoretically motivated and justified by DNS and experiment verified physical hypotheses, such as local dependence of the pressure hessian term on VGT, which were then subjected to physical constraints,  such as incompressibility, symmetries, etc. 
This modeling status quo, which seemed to be hard to improve, got a remarkable new boost from far away fields of Machine Learning (ML), Artificial Intelligence (AI), and Data Science.  Neural Networks (NNs) have gained significant attention due to their powerful capability in expressing complex nonlinear functions and the ease of parameter training through the Automatic Differentiation (AD) functionality provided by open-source ML packages. Empowered by these NNs, even though they are generally application-agnostic, researchers are now starting to use them to validate and improve scientific hypotheses and advance the modeling of complex physics problems.

To the best of our knowledge, there have been surprisingly few attempts to model Lagrangian dynamics using NNs or other ML methods, despite the rich information Lagrangian statistics can provide. This is in contrast to the surge of NN activity in recent years in the field of Eulerian modeling of turbulence. In particular, major efforts were devoted to the development of closure models for Reynolds Averaged Navier-Stokes (RANS) and Large Eddy Simulations (LES) using innovative NNs architectures, e.g. \citet{ling2016reynolds,maulik2019subgrid}. A very comprehensive overview of many contributions to this field can be found in the review paper by \citet{duraisamy2019turbulence}. To mention some of these contributions (which are mostly related to this manuscript): a Tensor Basis Neural Network (TBNN) embedding physical constraints, such as Galilean invariance and rotational invariance, into the closure model was developed in \citet{ling2016reynolds};  Physics-Informed Machine Learning (PIML) models infusing the NN with known physical constraints were developed in, e.g., \citep{2018PIML-LANL,wang2017physics}. Other than developing closure models for RANS and LES, researchers have been experimenting with novel ML approaches to learn turbulence dynamics. In this regards, and just to name a few, we mention \citet{mohan2019compressed}, where a Convolutional Long Short Term Memory (ConvLSTM) Neural Network was developed to learn spatio-temporal turbulence dynamics; studies of super-resolution allowing to reconstruct turbulence fields using under-resolved data \citet{fukami2019super}; and Neural Ordinary Differential Equation (Neural ODE) for turbulence forecasting \citet{portwood2019turbulence}.

In this manuscript, we combine the body of work in statistical hydrodynamics with ML tools to develop a physics-informed, interpretable model for Lagrangian dynamics of turbulence at both fully-resolved/unfiltered and coarse-grained/filtered levels.  \revision{Specifically, we build upon the original architecture of TBNN \citep{ling2016reynolds} and also extend TBNN with embedded physical constraints to model the non-local pressure Hessian and sub-filter dynamics within the Lagrangian description of VGT.}

The rest of the manuscript is organized as follows. Section \ref{sec:equation} provides technical background, e.g. discussing governing equations for Lagrangian dynamics of VGT.
PIML strategy for modeling closure terms, including physical interpretation of the underlying parameterization, are described in Section \ref{sec:piml}.  Section \ref{sec:dnsdata} is devoted to the description of the "ground-truth" DNS data used to train and test the models. The details of training and testing of the constructed PIML model are provided in Section \ref{sec:training}. Performance of the PIML models for unfiltered and also coarse-grained dynamics is assessed through a series of comprehensive tests in Sections \ref{sec:fullvgt} and \ref{sec:cvgt}, respectively. Finally, conclusions and path forward are discussed in Section \ref{sec:conclusion}.

\section{Governing equations\label{sec:equation}}

\subsection{Unfiltered Velocity Gradient Tensor}

The Lagrangian dynamics of the VGT in incompressible turbulence can be derived from the Navier-Stokes (NS) equations:
\begin{equation}
    \frac{\partial u_i}{\partial t}+u_k\frac{\partial u_i}{\partial x_k} = -\frac{\partial P}{\partial x_i} + \nu\frac{\partial^2 u_i}{\partial x_k \partial x_k},
\label{eqn:ns}
\end{equation}
where $u_i$, $P$, and $\nu$ denote velocity vector component, pressure, and kinematic viscosity, respectively. The velocity gradient tensor is defined as:
\begin{equation}
A_{ij} = \frac{\partial u_i}{\partial x_j}. 
\label{eqn:vgt}
\end{equation}
The Lagrangian dynamics of VGT and the incompressibility-enforcing Poisson equation can then be derived by applying spatial derivatives to Eq.~(\ref{eqn:ns}):
\begin{subequations}
\begin{eqnarray}
   \frac{d A_{ij}}{dt} =\frac{\partial A_{ij}}{\partial t} + u_k \frac{\partial A_{ij}}{\partial x_k} &=& -A_{ik}A_{kj} -\frac{\partial^2 P}{\partial x_i \partial x_j } + \nu \frac{\partial^2 A_{ij}}{\partial x_k \partial x_k}, \label{eqn:vgtdynamics} \\
   \frac{\partial^2 P}{\partial x_k \partial x_k}&=&-A_{mn}A_{nm}. \label{eqn:incompress}
\end{eqnarray}
\end{subequations}
Using Eq.~(\ref{eqn:incompress}), we then rearrange Eq.~(\ref{eqn:vgtdynamics}) into the following traceless form:
\begin{subequations}
\label{eqn:vgtdynam}
\begin{eqnarray}
   \frac{d A_{ij}}{dt} &=& E_{ij}+H_{ij}+T_{ij} \label{eqn:vgtdynamics2} \\
   E_{ij} &=& -(A_{ik}A_{kj}-\frac{1}{3}A_{mn}A_{nm}\delta_{ij}),\\
    H_{ij} &=& -\left(\frac{\partial^2 P}{\partial x_i \partial x_j }-\frac{1}{3}\frac{\partial^2 P}{\partial x_k \partial x_k} \delta_{ij}\right),\\
     T_{ij} &=& \nu \frac{\partial^2 A_{ij}}{\partial x_k \partial x_k},
\end{eqnarray}
\end{subequations}
where the matrices/tensors $\boldsymbol{E}$, $\boldsymbol{H}$,  and $\boldsymbol{T}$ are referred to as Restricted Euler (RE), non-local pressure Hessian, and viscous contributions to the Lagrangian dynamics, respectively. The RE equation is derived from Eq.~(\ref{eqn:vgtdynamics2}) by removing the pressure Hessian term and the viscous term. The resulting RE dynamics is closed. Obviously, the removal of the $\boldsymbol{E}$- and $\boldsymbol{H}$- contributions is not justified and closing VGT dynamics by modeling these terms,  i.e. expressing them via VGT, $\boldsymbol{A}$, or related local characteristics explicitly, is \revision{extremely challenging}.  

\subsection{Coarse-grained Velocity Gradient Tensor}

Not only Eq.~(\ref{eqn:vgtdynamics2}) is not closed, it is also spatially local,  therefore not showing (at least not explicitly) Lagrangian dynamics (and related statistics) of VGT at different spatial scales. In other words,  the local description is missing an important aspect of turbulence consisting in the understanding of how statistics of key characteristics and underlying phenomena, such as energy cascade and dynamics of particle separation, change with scale. 

To address this apparent lack of scale information in Eq.~(\ref{eqn:vgtdynamics2}), we apply a spatial filter (in this study we use a Gaussian filter for extracting the related quantities) to the NS Eq.~(\ref{eqn:ns}) and derive the transport equation for the so-called coarse-grained velocity gradient: 
\begin{equation}
    \frac{\partial \widetilde{u_i}}{\partial t}+\widetilde{u_k}\frac{\partial \widetilde{u_i}}{\partial x_k} = -\frac{\partial \tilde{P}}{\partial x_i} + \nu\frac{\partial^2 \widetilde{u_i}}{\partial x_k \partial x_k} - \frac{\partial \tau_{ik}}{\partial x_k},
\label{eqn:nsfilter}
\end{equation}
where the tilde sign, e.g. $\tilde{u}$, marks filtered variables. The filter is characterized by its scale and typically selected to be within the inertial range of turbulence, which is bounded from above by the energy-containing scale and from below by the viscous (Kolmogorov) scale. The sub-filter contributions,  i.e. these associated with scales smaller than the filtering scale, are marked by apostrophe, e.g. $u'$,  so that $u=\tilde{u}+u'$. $\tau$, entering the filtered Eq.~(\ref{eqn:nsfilter}), is the so-called sub-filter stress.  The sub-filter stress contribution on the right-hand side of Eq.~(\ref{eqn:nsfilter}) originates from the nonlinearity of the original NS Eq.~(\ref{eqn:ns}) and makes the resulting filtered equations not closed.

One can also apply the filtering procedure to Eq.~(\ref{eqn:vgtdynamics}), therefore arriving at the following equation governing Lagrangian dynamics of the coarse-grained VGT: 
\begin{subequations}
\begin{eqnarray}
   \frac{d M_{ij}}{dt} &=& \widetilde{E_{ij}}+\widetilde{H_{ij}} + \widetilde{T_{ij}} + \Pi_{ij}\label{eqn:vgtdynamics3} \\
   \widetilde{E_{ij}} &=& -(M_{ik}M_{kj}-\frac{1}{3}M_{mn}M_{nm}\delta_{ij}),\\
    \widetilde{H_{ij}} &=& -(\frac{\partial^2 \widetilde{P}}{\partial x_i \partial  x_j }-\frac{1}{3}\frac{\partial^2 \widetilde{P}}{\partial x_k \partial x_k} \delta_{ij}),\\
     \widetilde{T_{ij}} &=& \nu \frac{\partial^2 M_{ij}}{\partial x_k \partial x_k}\\
     \Pi_{ij} &=&\Pi_{ij}^{r}+\Pi_{ij}^{s}= -(\frac{\partial^2 \tau_{ik}}{\partial x_j \partial x_k} -\frac{1}{3}\frac{\partial^2 \tau_{ik}}{\partial x_j \partial x_k} \delta_{ij}) + u'_k\frac{\partial M_{ij}}{\partial x_k},\\
     \Pi_{ij}^{r} &=& -(\frac{\partial^2 \tau_{ik}}{\partial x_j \partial x_k} -\frac{1}{3}\frac{\partial^2 \tau_{ik}}{\partial x_j \partial x_k} \delta_{ij}), \\
     \Pi_{ij}^{s} &=& u'_k\frac{\partial M_{ij}}{\partial x_k},
\end{eqnarray}
\end{subequations}
where $\boldsymbol{M}$ is our notation for the filtered component of $\boldsymbol{A}$. Notice the appearance of a new (if compared with the original, unfiltered equation) sub-grid contribution, $\boldsymbol{\Pi}$, on the right-hand side of Eq.~(\ref{eqn:vgtdynamics3}). Furthermore, taking the divergence of the filtered version of the momentum equations and using commutation of the filtering operation with respect to spatial derivative, one arrives at the following filtered version of the Poisson equation expressing filtered pressure via $\boldsymbol{M}$:
\begin{eqnarray}
   \frac{\partial^2 \widetilde{P}}{\partial x_k \partial x_k}&=&-M_{mn}M_{nm}-\frac{\partial^2 \tau_{ik}}{\partial x_i \partial x_k} \label{eqn:incompress2}
\end{eqnarray}
This equation has additional terms on the right-hand side compared to Eq.~(\ref{eqn:incompress}). To make the Lagrangian dynamics of the coarse-grained VGT closed, i.e. expressed via $\boldsymbol{M}$, one obviously needs to model functional dependence of the pressure Hessian, viscous, and sub-filter contributions on $\boldsymbol{M}$.

\section{PIML strategy\label{sec:piml}}

In this section, we describe our approach to modeling the generally unclosed unfiltered and coarse-grained Lagrangian dynamics, described by Eqs.~(\ref{eqn:vgtdynamics2}) and (\ref{eqn:vgtdynamics3}), via a combination of NNs and physical constraints.

\subsection{Physical Foundation of PIML Model}

\subsubsection{Non-local Pressure Hessian}

Formal solution for non-local pressure Hessian, expressed as a convolution of the second invariant of the VGT, can be obtained by solving equation (\ref{eqn:incompress}) with the Green's function of the Laplacian operator \citep{ohkitani1995nonlocal}: 
\begin{subequations}
\begin{eqnarray}
\label{eqn:hij}
H_{ij}(\boldsymbol{x}) &=& \int\int\int\frac{\delta_{ij}-\hat{r_i}\hat{r_j}}{2\pi r^3}Q(\boldsymbol{x}+\boldsymbol{r})d\boldsymbol{r},\\
Q&=& -\frac{1}{2} A_{mn}A_{nm}, \label{eqn:q}
\end{eqnarray}
\label{eqn:incompress3}
\end{subequations}
\noindent where $\boldsymbol{r}$ indicates the displacement vector with respect to the point of interest $\boldsymbol{x}$, and  $\hat{\boldsymbol{r}} = \boldsymbol{r} / r$ is the unit displacement vector. Eqs.~(\ref{eqn:incompress3}) show that the non-local pressure Hessian depends on the spatial distribution of the second invariant of the velocity gradient tensor, $Q(\boldsymbol{x}+\boldsymbol{r})$. 

In this work, we model the non-local pressure Hessian term by a surrogate parameterized expression for the spatial distribution of the second invariant $Q$. We substitute $Q(\boldsymbol{x}+\boldsymbol{r})$, similar to the approach used in \citet{lawson2015velocity}, by a function of  $\boldsymbol{A}$. Assuming locality of the relation, which is admittedly a bold hypothesis, the most general form of $Q(\boldsymbol{x}+\boldsymbol{r})$ is the Taylor series expansion:
\begin{equation}
Q(\boldsymbol{x}+\boldsymbol{r}) = a(\boldsymbol{r}) + b_{ij}(\boldsymbol{r})A_{ij} + c_{ijkl}(\boldsymbol{r})A_{ij}A_{kl} + ...
\label{eqn:taylor}
\end{equation}
Next, and following \citet{lawson2015velocity}, we enforce in  Eq.~(\ref{eqn:taylor}) rotational invariance, i.e.  $Q(\boldsymbol{x}+\boldsymbol{r}| \boldsymbol{A}) =  Q(\boldsymbol{x}+\boldsymbol{Vr}| \boldsymbol{VAV^T})$,  then arrive at:
\begin{eqnarray}
Q(\boldsymbol{x}+\boldsymbol{r}) &=& b_1 \tau_0^{-2} +b_2 (\hat{\boldsymbol{r}}^T\boldsymbol{A}\hat{\boldsymbol{r}}) + b_3Tr(\boldsymbol{A}^2) + b_4Tr(\boldsymbol{A}\boldsymbol{A}^T) + b_5 (\hat{\boldsymbol{r}}^T\boldsymbol{A}^2\hat{\boldsymbol{r}}) \nonumber \\
&&+b_6 (\hat{\boldsymbol{r}}^T\boldsymbol{AA}^T\hat{\boldsymbol{r}}) +b_7 (\hat{\boldsymbol{r}}^T\boldsymbol{A}^T\boldsymbol{A}\hat{\boldsymbol{r}}) +b_8 (\hat{\boldsymbol{r}}^T\boldsymbol{A}\hat{\boldsymbol{r}})^2+...
\label{eqn:taylor2}
\end{eqnarray}
where $\tau_0$ is the reference timescale and $b_i$ for $i=1,\cdots,$ are scalar functions of $r$, assumed related to the two-point correlation functions of $Q$ with different moments of VGT.  In the models analyzed so far, the expansion in Eq.~(\ref{eqn:taylor2}) is usually truncated at the second level,  see e.g. \citep{wilczek2014pressure,lawson2015velocity}.  In this study, we take advantage of the ML approach and consider full expansion, then extracting functional forms of the coefficients through training.  

Equation (\ref{eqn:taylor2}) could be combined  with equation (\ref{eqn:incompress3}) to perform the integration over $\boldsymbol{r}$. We use the strain rate tensor $\boldsymbol{S}$ and rotation tensor $\boldsymbol{W}$ to denote the symmetric and skew-symmetric parts of the VGT, where $\boldsymbol{A} = \boldsymbol{S} +\boldsymbol{W}$ and $\boldsymbol{A}^T = \boldsymbol{S} -\boldsymbol{W}$. Our final model for the non-local pressure Hessian becomes:
\begin{eqnarray}
\boldsymbol{H} = \sum_{m,n=0}^{\infty} a_{mn} \boldsymbol{S}^m\boldsymbol{W}^n.
\label{eqn:phessian1}
\end{eqnarray}
This expression coincides with the model for effective viscosity proposed in \citet{pope1975more} to represent the anisotropy tensor of Reynolds stresses in the RANS Eulerian approach. The coefficients in Eq.~(\ref{eqn:phessian1}) represent weighted integrals of the two-point correlation function of $Q$ with different moments of VGT \citep{lawson2015velocity}. Originally, i.e. in \citet{pope1975more},  the equation was further transformed utilizing the symmetric integrity bases and invariants of the  $\boldsymbol{S}$ and $\boldsymbol{R}$ tensors and then using the Caley-Hamilton theorem to reduce the high-order products of the tensor to linear combinations of the integrity bases: 
\begin{equation}
\boldsymbol{H} = \sum_{n=1}^{10} g_s^{(n)}(\lambda_1,...,\lambda_5)\boldsymbol{T}_s^{(n)}, \label{eqn:phessian2}
\end{equation}
where $g_s^{(n)}$ are scalar functions of the invariants. In total, there are 5 independent invariants and 10 integrity bases \citep{pope1975more}:
\begin{align}
\lambda_1&=Tr(\boldsymbol{S}^2), &\lambda_2 =& Tr(\boldsymbol{W}^2), &\lambda_3 = Tr(\boldsymbol{S}^3), \nonumber \\
\lambda_4&=Tr(\boldsymbol{W}^2\boldsymbol{S}),  &\lambda_5 =& Tr(\boldsymbol{W}^2\boldsymbol{S}^2)& 
\label{eqn:invariant}
\end{align}
\begin{align}
\boldsymbol{T}_s^{(1)}&=\boldsymbol{S}, & \boldsymbol{T}_s^{(2)} =&\boldsymbol{SW-WS}, & \nonumber \\
\boldsymbol{T}_s^{(3)} &=\boldsymbol{S}^2-\frac{1}{3}\boldsymbol{I}Tr(\boldsymbol{S}^2), &\boldsymbol{T}_s^{(4)} =&\boldsymbol{W}^2-\frac{1}{3}\boldsymbol{I}Tr(\boldsymbol{W}^2),  &\nonumber \\
\boldsymbol{T}_s^{(5)} &=\boldsymbol{WS}^2-\boldsymbol{S}^2\boldsymbol{W},& \boldsymbol{T}_s^{(6)} =&\boldsymbol{W}^2\boldsymbol{S}+\boldsymbol{SW}^2-\frac{2}{3}Tr(\boldsymbol{SW}^2), & \nonumber\\
\boldsymbol{T}_s^{(7)}& =\boldsymbol{WSW}^2-\boldsymbol{W}^2\boldsymbol{SW}, & \boldsymbol{T}_s^{(8)} =&\boldsymbol{SWS}^2 - \boldsymbol{S}^2\boldsymbol{WS},& \nonumber\\
\boldsymbol{T}_s^{(9)} &=\boldsymbol{W}^2\boldsymbol{S}^2+\boldsymbol{S}^2 \boldsymbol{W}^2-\frac{2}{3}\boldsymbol{I}Tr(\boldsymbol{S}^2\boldsymbol{W}^2), & \boldsymbol{T}_s^{(10)} =&\boldsymbol{WS}^2\boldsymbol{W}^2 - \boldsymbol{W}^2\boldsymbol{S}^2\boldsymbol{W} &
\label{eqn:bases}
\end{align}
We emphasize that in all the previous studies known to the authors, 
the functional form of $g^{(n)}$ was greatly simplified and the integrity bases considered were typically limited to second-order \citep{wilczek2014pressure,lawson2015velocity}, i.e. the first four bases $\boldsymbol{T}_s^{(1)}, ..., \boldsymbol{T}_s^{(4)}$. Then, $g_s^{(n)}(\lambda_1,...\lambda_5)$ was either derived from a modeled two-point correlation function or reconstructed from experimental or numerical data. In this work, we use a ML model, or more specifically a NN, to learn the functional form of $g_s^{(n)}$.

\subsubsection{Sub-Filter Contribution}
\label{sec:subfilter}

The sub-filter contribution $\boldsymbol{\Pi}$ in the coarse-grained dynamics of the VGT can be divided into two parts: sub-filter stress $\boldsymbol{\Pi}^r$ and sub-filter fluctuations $\boldsymbol{\Pi}^s$. We take different approaches to model the two terms. 

The sub-filter stress contribution is non-local and asymmetric in nature, and we use an approach similar to the one proposed in \citet{pope1975more} to model the term. We expand the integrity bases to account for the skew-symmetric part and thus consider 
\begin{equation}
\boldsymbol{\Pi}^{r} = \sum_{n=1}^{10} g_s^{(n)}(\lambda_1,...,\lambda_5)\boldsymbol{T}_s^{(n)} + \sum_{n=1}^{6} g_a^{(n)}(\lambda_1,...,\lambda_5)\boldsymbol{T}_a^{(n)}.
\label{eqn:rstress}
\end{equation}
There are 6 skew-symmetric bases and they can be derived following \citet{zheng1993representations}:
\begin{align}
\boldsymbol{T}_a^{(1)}&=\boldsymbol{W}, & \boldsymbol{T}_a^{(2)} =&\boldsymbol{SW+WS}, & \nonumber \\
\boldsymbol{T}_a^{(3)} &=\boldsymbol{S}^2\boldsymbol{W}+\boldsymbol{WS}^2, &\boldsymbol{T}_a^{(4)} =&\boldsymbol{W}^2\boldsymbol{S}-\boldsymbol{SW}^2,  &\nonumber \\
\boldsymbol{T}_a^{(5)} &=\boldsymbol{W}^2\boldsymbol{S}^2-\boldsymbol{S}^2\boldsymbol{W}^2,& \boldsymbol{T}_a^{(6)} =&\boldsymbol{SW}^2\boldsymbol{S}^2-\boldsymbol{S}^2\boldsymbol{W}^2\boldsymbol{S}. &
\label{eqn:bases2}
\end{align}

The sub-filter fluctuations come from the spatiotemporal scales that are  smaller than the filter size. 
Specifically,  the sub-filter contribution to the coarse-grained dynamics is associated with the effects of backscatering, i.e. transfer of energy from scales smaller than the filter scale upscales, which goes against the average transfer of energy towards smaller scales. We choose to model these backscattering contributions by  a stochastic Langevin process:
\begin{equation}
\Pi^s_{ij} = C_{ijkl}dW_{kl}    
\end{equation}
where d$\boldsymbol{W}$ stands for the Wiener process. The co-variance matrix $\boldsymbol{C}$ is fixed and expected to be learned from the "ground-truth" data.


\subsection{Neural Network Architecture}

It is well known since at least the 90s of last century, see e.g. \citep{hornik1991approximation},  that NNs are capable of representing an arbitrary smooth function. This theoretical concept got a tremendous boost empirically during the last decade with the so-called Deep Learning, which made the approach practically feasible, and moreover a powerful tool applicable to a great variety of applications. In this work, we take advantage of the modern Deep Learning approach to model different contributions to the Lagrangian dynamics of VGT. \revision{We also test the benefits of injecting known physics/constraints into NNs, by comparing the proposed Physics-Informed Machine Learning (PIML) strategy to an application ignorant NN approach, which we refer to as Physics-Blind (PB). }

\subsubsection{Extended Tensor Basis Neural Network}


To learn the functions contributing to Eq.~(\ref{eqn:phessian1}), we follow the approach of \citet{ling2016reynolds}, where a Tensor Basis Neural Network (TBNN) was developed. TBNN maps functions of invariants and integrity bases to another target tensor. The method was used in \citet{ling2016reynolds} and in a number of follow-up papers to develop ML models for RANS and LES. In this work, we use an \revision{extended TBNN approach to learn our physics-informed Lagrangian dynamics of VGT.}

\begin{figure}[hbt]
	\includegraphics[width=3in]{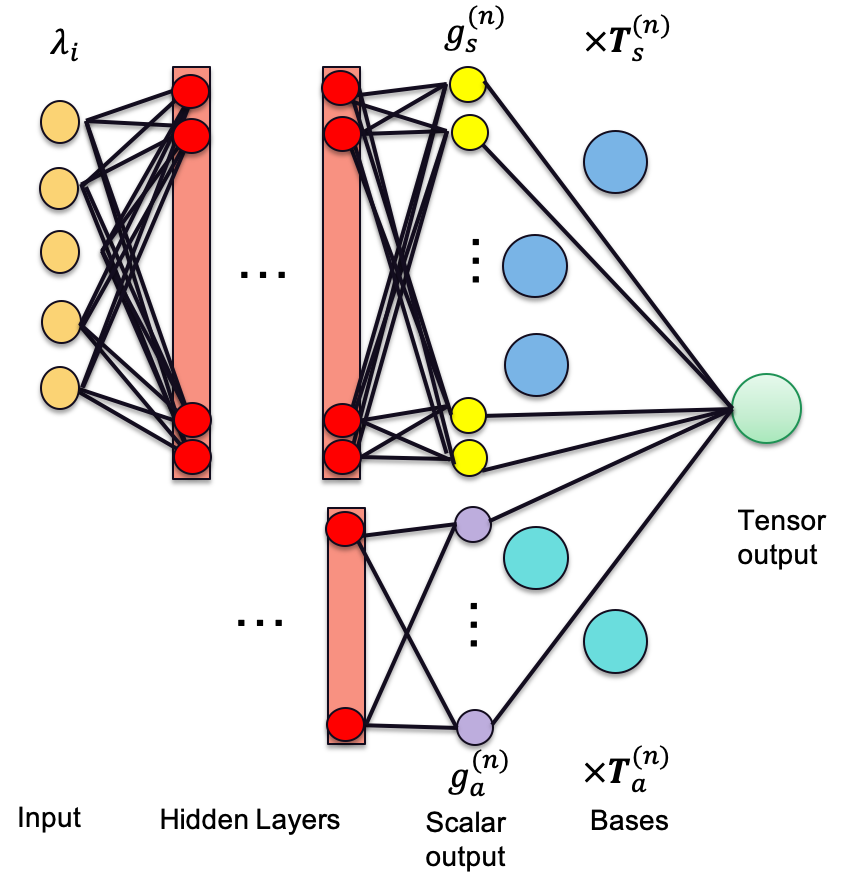}
	\caption{Architecture of the extended Tensor Basis Neural Network used in this work, which is based on the TBNN structure used in \citet{ling2016reynolds} and extended to include skew-symmetric tensor bases.}
	\label{fig:tbnn}
\end{figure}

The architecture of the extended TBNN we use in this study is shown in Fig.~\ref{fig:tbnn}. Inputs of the multi-layer Deep NN are the 5 invariants $\lambda_i$ introduced above. The input layer is connected to a series of hidden layers via the \revision{Rectified Linear Unit or} ReLU activation function, \revision{which is used to introduce nonlinearities into the ML model}. The first output layer is the scalar output with 2 branches, each representing scalar functions $g_s^{(n)}$ and $g_a^{(n)}$. The scalar output is then combined with the corresponding symmetric and skew-symmetric integrity bases to produce the final tensor output. The extended TBNN expands the tensor bases where the output tensor is residing on. While retaining the classical features of TBNN, such as Galilean invariance and rotational invariance,  we also guarantee in this architecture enforcement of the incompressibility constraint by modeling the deviatoric/traceless part of the dynamics.

The choice of the hyper-parameters of the Deep NN, such as the number of layers and number of nodes per layer, is proven to be critical. The NN needs to be deep enough to represent a sufficiently wide class of functions,  however also not too deep to avoid over-fitting. We have established, by trial-and-error, that a NN with 8 layers and 40 nodes per layer in the hidden layer achieves a good balance. The scalar output layer has either 10 or 16 nodes depending on the types of integrity bases used. The last layer is a "physical" layer with no trainable parameters. The training is performed using the "Adam" optimization method \citep{kingma2014adam} with a standard (quadratic) loss function. We use a varying learning rate, which decreases from $10^{-3}$ to $10^{-6}$ throughout the process of training. 

\subsubsection{Physics-Blind Deep Neural Network}
\revision{In order to demonstrate advantages of the proposed PIML model, we construct a naive Physics-Blind model which does not assume any prior physics input. A natural choice of the NN structure for the PB model is the out-of-the-box Deep Neural Network, which is a multi-layer, densely connected, feed-forward neural network. This Deep Neural Network gets 9 components of the non-dimensionalized velocity gradient tensor as input and it outputs 9 components of the non-local pressure Hessian. In terms of the hyper-parameters, we choose to use a NN with 8 layers and 40 nodes within each layer, resulting in the number of parameters comparable to the number of trainable parameters within the PIML model. The learning rates of the PB model and the PIML model are the same.}

\section{"Ground-Truth" Data}
\label{sec:dnsdata}

The "ground-truth" data is generated from the Eulerian DNS solution of the incompressible Navier-Stokes equations. Isotropic turbulence is generated on a $1024^3$ grid using the pseudo-spectral method. A large-scale linear forcing term is applied to prevent turbulence from decaying. Time advancement is achieved through Adam-Bashforth-Moulton method. The Taylor Reynolds number when the turbulence reaches a statistically steady state is approximately $250$. See \citet{petersen2010forcing,daniel2018reaction} for more details on the numerical method. The coarse-grained statistics are obtained by applying a Gaussian filter to the Eulerian data with a chosen filter size $l_{\Delta}$.

Only Eulerian data are required for the training purpose. In this work, we have sampled up to 4 million data points randomly spanned over space and time, but no improvement is observed when the training set is larger than 0.5 million. \revision{Then, turbulence statistics of interest, including statistics of the coarse-grained VGT, pressure Hessian, viscous term, and sub-filter term, are calculated using 6th-order Lagrange interpolation from the Eulerian grid to the random locations. The data are then non-dimensionalized using the corresponding time scales of the resolved VGT (i.e. Kolmogorov time scale for unfiltered VGT). The invariants and the tensor bases, which are inputs of the extended TBNN, are then calculated using the dimensionless VGT.} The train-to-test split ratio is chosen to be 80\%:20\%.

The Lagrangian data following trajectories of fluid particles are also evaluated and used for diagnostics. The fluid particles are treated as non-inertial particles and their velocity is assumed to equal the local flow velocity. The velocity of the fluid particle is interpolated from the Eulerian grid using cubic splines \citep{yeung1988algorithm}. Trajectories of the fluid particles and the temporal evolution of various statistics are recorded and stored to be used later for training and validation.

\section{Training and Testing of the ML Models}
\label{sec:training}
\begin{figure}[hbt]
	\includegraphics[width=6in,height=2.5in]{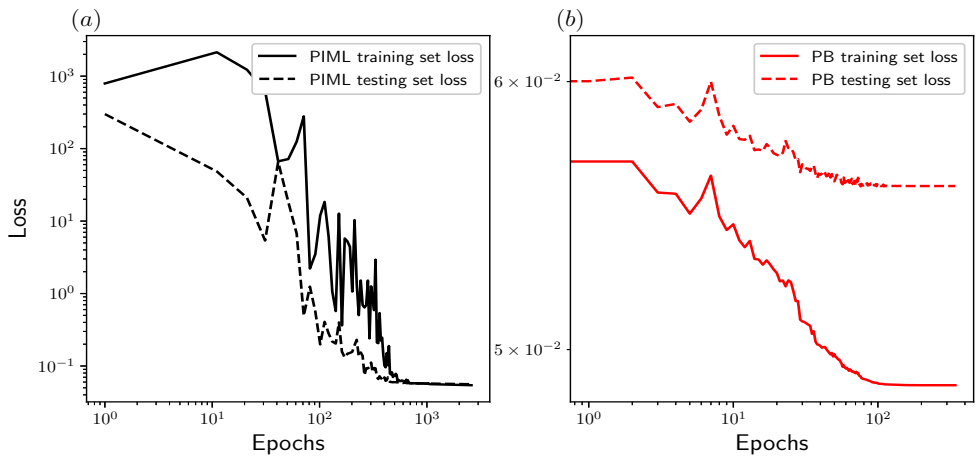}
	\caption{\revision{The decay of both the training and testing loss for the (\textit{a}) PIML model and (\textit{b}) PB model.}}
	\label{fig:loss}
\end{figure}

\revision{We implement and train the PIML model and the PB model using TensorFlow \citep{tensorflow2015-whitepaper} and Keras \citep{chollet2015keras} open-source ML library. The parameters (weights and biases of the NN layers) in both models are randomly initialized using the Glorot normal initialization method. The dataset is randomly divided into mini-batches of size 4096, therefore, each training epoch takes about 122 iterations. The parameters of the NN are updated by minimizing the quadratic loss function:

\begin{equation}
    \label{eqn:loss}
    Loss = \frac{1}{n_{train}} \sum_{i=1}^{n_{train}} \Vert \boldsymbol{H}_{i,model} - \boldsymbol{H}_{i,GT} \Vert_F^2,
\end{equation}
\noindent using the Adam optimization method. Here in Eq.~(\ref{eqn:loss}), $\Vert \cdot \Vert _F$ denotes the Frobenius norm of tensors. The initial learning rate is set to $10^{-3}$ and it is then reduced gradually to $10^{-6}$ throughout the training process. Training is terminated when both training and testing losses are saturrated. Fig. \ref{fig:loss} shows the training and testing losses for both PIML and PB models as a function of the number of epochs. We observe that the initial loss for the PIML model is much larger than one observed in the PB model. This is due to the fact that the high-order tensor basis is highly intermittent, and therefore random  initialization of the parameters results in a large loss. Since the initial loss of the PIML model is orders of magnitude larger than one in the PB model, it takes more epochs for the PIML model to fully converge. At the final training epoch, the PIML model reaches a similar training/testing loss of $4.80\times 10^{-2}$/$4.90\times 10^{-2}$. On the other hand, the final testing loss is $5.58\times 10^{-2}$ for the PB model, 
that is larger than the training loss of PIML.
This shows that injecting the known physical constraints into the NN structure can provide physics-based regularization for training of the ML models.}

\begin{figure}[hbt]
	\includegraphics[width=6in,height=2.5in]{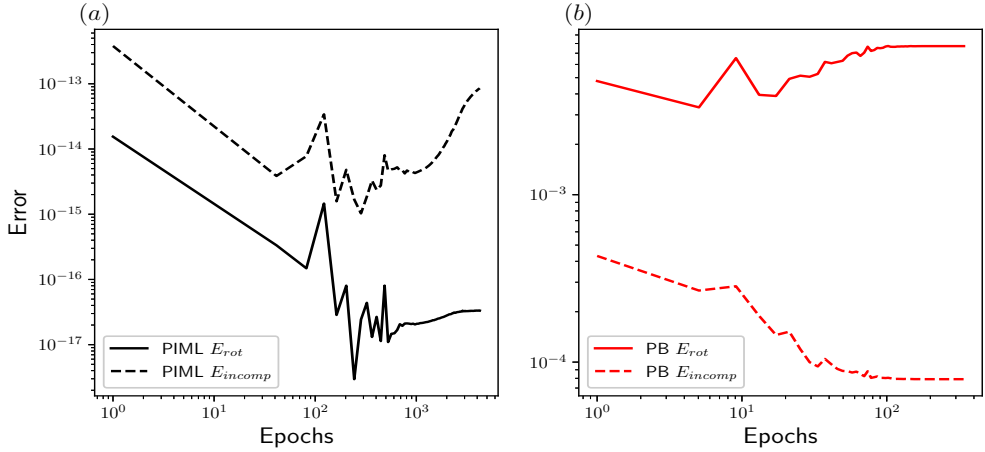}
	\caption{\revision{The testing error of the incompressibility constraint and rotational invariance of the (\textit{a}) PIML and (\textit{b}) PB models. }}
	\label{fig:invariances}
\end{figure}

\revision{Another advantage of the physics-informed approach is that the physical constraints and invariance can be automatically preserved by the build-in NN structure. On the other hand, when working with the PB model, we blindly rely on the NN to discover the underlying physical constraints and invariance. In Fig. \ref{fig:invariances}, we provide comparison of the testing errors of the incompressibility constraints ($E_{incomp}$) and rotational invariance ($E_{rot}$) from the PIML and PB models. The errors are calculated using the following formulas:
\begin{subequations}
\begin{eqnarray}
\label{eqn:invariance}
E_{incomp} &=& \frac{1}{n_{test}} \sum_{i=1}^{n_{test}} \vert Tr(\boldsymbol{H})\vert, \\
E_{rot} &=& \frac{1}{n_{test}} \sum_{i=1}^{n_{test}} \Vert \boldsymbol{\Omega} f(\boldsymbol{A}) - f(\boldsymbol{\Omega} \boldsymbol{A})
\Vert_{F}^2,
\end{eqnarray}
\end{subequations}
\noindent where $\boldsymbol{\Omega}$ denotes a random rotation matrix and $f$ denotes the ML models. As expected, the PIML model conserves both incompressibility constraints and rotational invariance throughout the training and respective errors are around the machine precision. On the other hand, the PB model is unable to fully capture the underlying invariance.
}

\section{Unfiltered Velocity Gradient Tensor\label{sec:fullvgt}}

The PIML model described in Section \ref{sec:piml} is trained using fully-resolved data for the VGT and the pressure Hessian statistics. In this section, the performance of the trained PIML Lagrangian model is analyzed from a number of perspectives. The unclosed non-local pressure Hessian is in the prime focus of the analysis. A stochastic ODE is then constructed using the newly learned model to check if the statistics is reproducible.

\subsection{{\em A Priori} Results}

 Aside from the Frobenius norm of the tensor, which is used as the loss function in training, an important diagnostic of the predicted pressure Hessian is associated with the alignment of its eigenvectors and magnitudes of its eigenvalues with respect to characteristics extracted from the "ground-truth" data. Directions of the eigenvectors are known to play an important role in the dynamics of VGT, which are related to flow topology, vortex stretching mechanism, dissipation, etc. In this section, the PIML model of the non-local pressure Hessian is developed using the extended TBNN.  It is then compared with a number of prior attempts to model the non-local pressure Hessian. Specifically,  we juxtapose our model to the empirical models developed in \citet{wilczek2014pressure,lawson2015velocity,johnson2016closure}, which rely on the aforementioned arguments for the pressure Hessian and sub-grid terms locality, but do not use a NN to represent functional dependence of the modeled terms on the VGT. To emphasize the role of higher-order functions in our local modeling, we also introduce hierarchy of models by constructing two simplified ML models: Model A, for which $g_s^{(n)}$ are set to constants (rather than functions of $\lambda_i$), and Model B, for which $g_s^{(n)}$ are set to linear functions of $\lambda_i$.

\begin{figure}[hbt]
	\includegraphics[width=4in]{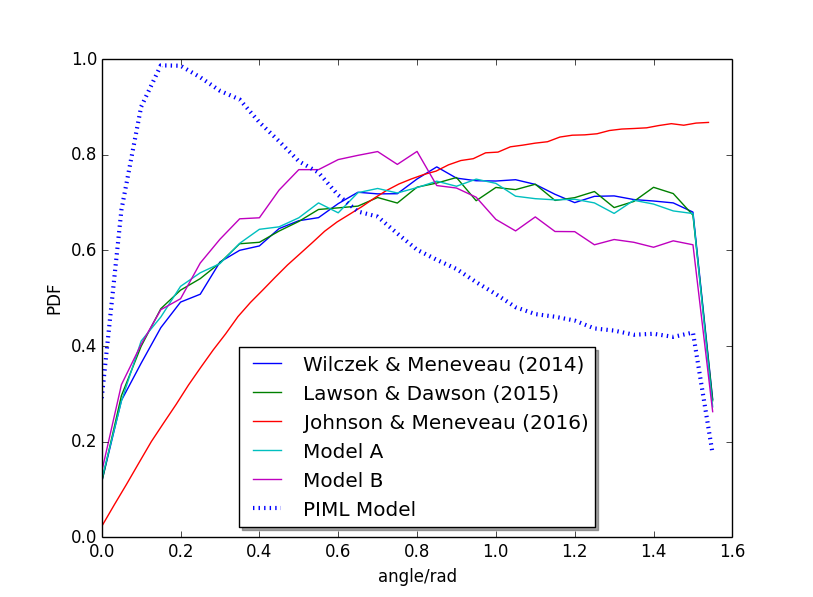}
	\caption{PDFs of the angle (in radians)  between the predicted $e_1$ eigenvector and DNS data. \revision{The PDF peaks at 0 when the predicted $e_1$ eigenvector is aligned with the DNS data and it peaks at $\pi/2$ in the case of mis-alignment.}}
	\label{fig:pdfangle1}
\end{figure}

Symmetries of the non-local pressure Hessian allow us to reduce the discussion to an analysis of its three distinct real eigenvalues (denoted as $e_1$ $<$ $e_2$ $<$ $e_3$) and respective eigenvectors. 
Fig.~\ref{fig:pdfangle1} shows the PDF of the angle between predicted and "ground-truth" directions of the $e_1$ eigenvector, corresponding to the most negative eigenvalue. If the PDF peaks at $(\pi/2)$ rad, it indicates that the model predictions are mostly misaligned with the "ground-truth" data. If the PDF peaks at $0$ rad, the model eigenvector is likely to be aligned with the ground-truth data. Perfect alignment with a single-valued PDF of $0$ rad is impossible because of the approximate nature of the empirical model (we approximate a clearly non-local problem with a local model). Fig.~\ref{fig:pdfangle1} shows that the only model that can capture the alignment is our PIML model. Second in performance is the simplified ML model B (with $g_s^{(n)}$ assumed to be linear in $\lambda_i$). Even though the performance of the linearized PIML models (both ML model A and ML model B) are somehow better than that of other non-ML models, the improvement is limited. We relate the poor performance of the non-ML models in the alignment test to the fact that the models ignore high-order functions of the invariants and bases. Simplified ML models A and B improve the representation of the functional space; however, in only a limited way as the coefficients are restricted to being constants or linear functions. The ability of our basic PIML model to learn sufficiently many higher-order functions makes it significantly better in performance than other models. The limitations of the non-ML based models were foreshadowed in \citet{lawson2015velocity}.

The eigenvalues of the non-local pressure Hessian, known to play an important role in explaining physical details of the energy cascade, type of the fluid element deformation, vorticity-strain alignment, and vortex stretching are tested next.  
By construction, the trace of the non-local pressure Hessian tensor is zero, so that the sum of the eigenvalues is also zero, $e_1+e_2+e_3=0$. We denote these eigenvalues such as $e_3$ is largest (mostly positive) eigenvalue, $e_1$, is the smallest (mostly negative) eigenvalue, and $e_2$ is the intermediate eigenvalue and can be either negative or positive. 
Table \ref{tab:eigval} shows correlation coefficients between model prediction and DNS (ground truth) for the three eigenvalues. Evidently, the PIML model shows a significant improvement over all other models for all the eigenvalues, therefore giving a better prediction of the dynamics. Another observation is that the correlation coefficients corresponding to $e_1$ and $e_3$ are generally much higher than those for $e_2$. We relate it to the fact that by construction $e_2$ is smaller in magnitude, and it can also be positive or negative,  therefore making it harder to predict its mean value accurately.  

\begin{table*}
\caption{\label{tab:eigval}%
Correlation coefficients of the eigenvalues.
}
\begin{ruledtabular}
\begin{tabular}{cccc}
\textrm{Models}&
\textrm{corr($e_{1,model},e_{1,DNS}$)}&
\textrm{corr($e_{2,model},e_{2,DNS}$)} & 
\textrm{corr($e_{3,model},e_{3,DNS}$)} \\
\colrule
\citet{wilczek2014pressure} & 0.69 & 0.11 & 0.71\\
\citet{lawson2015velocity} & 0.69 & 0.04 & 0.72\\
\citet{johnson2016closure} & 0.13 & 0.06 & 0.12\\
Model A & 0.70 & -0.07 & 0.73\\
Model B   & 0.61 & 0.09 & 0.67\\
PIML model & 0.95 & 0.79 & 0.94\\

\end{tabular}
\end{ruledtabular}
\end{table*}

\begin{figure}[hbt]
	\includegraphics[width=6in]{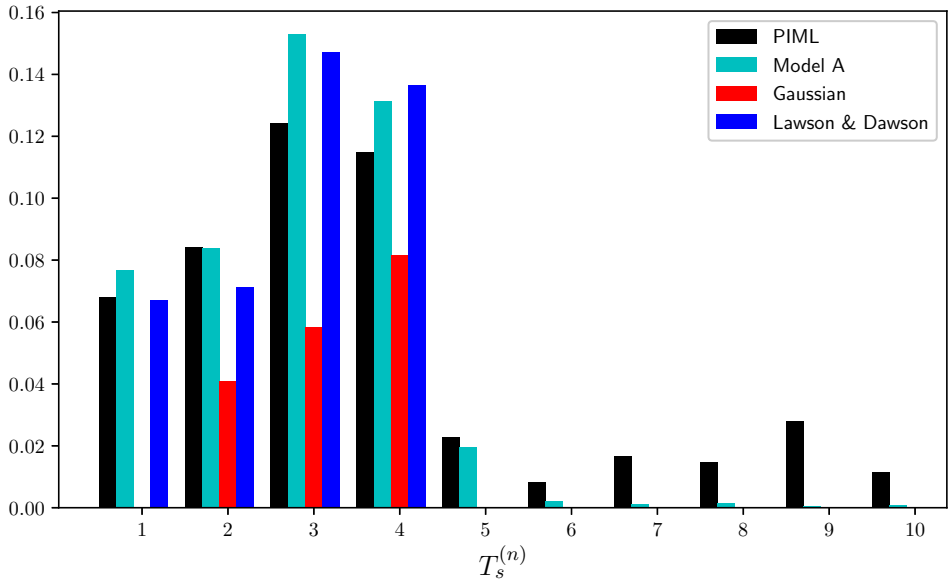}
	\caption{\revision{Contributions of the different tensor bases to the prediction of non-local pressure Hessian for PIML model, ML model A, Gaussian closure \citep{wilczek2014pressure}, and \citet{lawson2015velocity}.}}
	\label{fig:hist}
\end{figure}

\revision{The advantage of the PIML model based on TBNN is that, unlike in the case of the black-box Physics-Blind model, it allows to interpret the trained model by extracting the scalar coefficients and compare them with physics-based models. The general functional form of previous physics-based models used in the past \citep{wilczek2014pressure,lawson2015velocity,leppin2020capturing} can be written as:
\begin{equation}
\label{eqn:physicsform}
\boldsymbol{H} = \delta \boldsymbol{S} + \gamma \l (\boldsymbol{SW-WS} \r ) + \alpha \l ( \boldsymbol{S}^2-\frac{1}{3}\boldsymbol{I}Tr(\boldsymbol{S}^2) \r ) + \beta \l ( \boldsymbol{W}^2-\frac{1}{3}\boldsymbol{I}Tr(\boldsymbol{W}^2) \r ).
\end{equation}
\noindent The main tuning parameters for the model are $\delta$, $\gamma$, $\alpha$, and $\beta$, which can be viewed as the scalar coefficients $g_s^{(1)}$ to $g_s^{(4)}$. In Fig. \ref{fig:pdfparam}, we compare the PDFs of the 4 scalar coefficients extracted from the PIML model and the ML model A, which are computed using the testing set with the corresponding tuning parameters taken from previous studies, i.e. based on the Gaussian closure \citep{wilczek2014pressure} and on the second-order basis expansion \citep{lawson2015velocity}. Note that model A is based on the same TBNN structure, however with 10 scalar functions $g_s^{(n)}$ set constant. We observe that the learned scalar coefficients in the model A are very close to the parameters in \citet{lawson2015velocity}, but the coefficients are very different from those derived using the Gaussian closure model. This is because the parameters in \citet{lawson2015velocity} are also obtained by matching with DNS data using second-order basis expansion. In the case of the PIML model, the predicted scalar coefficients are no longer constant --  they are nonlinear functions of the invariants.  Fig. \ref{fig:pdfparam} shows that by expanding the functional space of the scalar coefficients using a NN, the model is able to capture non-trivial nonlinear dependence on the invariants originating from the projection of high-order polynomials to the low-order bases. This non-trivial nonlinear dependence shifts the peak values of the PDFs away from the constant coefficients in model A. We also note that the PDFs of the learned scalar coefficients are generally narrow in the center but also showing wide tails. This is likely due to the high intermittency of the high-order invariants. 

\begin{figure}[hbt]
	\includegraphics[width=6in]{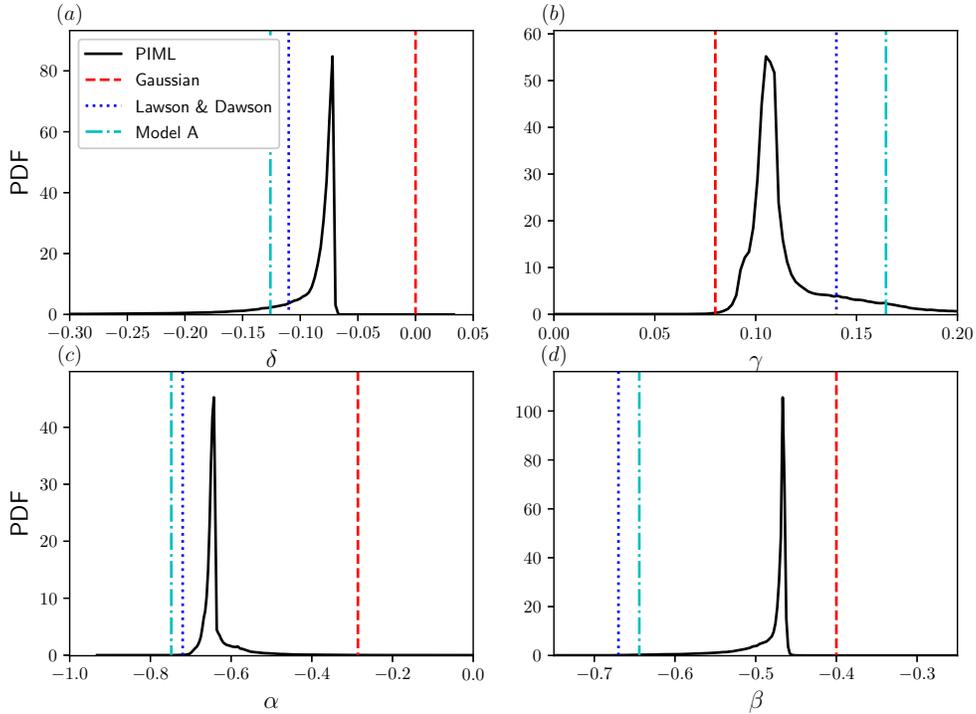}
	\caption{\revision{PDFs of the coefficients of the tensor bases from PIML model, ML model A, Gaussian closure \citep{wilczek2014pressure} and \citet{lawson2015velocity}.}}
	\label{fig:pdfparam}
\end{figure}

We then compare contributions from different integrity bases using the average Frobenius norm of $g_s^{(n)}\boldsymbol{T}^{(n)}$ for different models. As shown in Fig. \ref{fig:hist}, the Frobenius norm of contributions from the first 4 integrity bases are comparable across the PIML model, model A, and \citet{lawson2015velocity}, and they are also larger than those seen in the Gaussian closure model. For high-order bases, model A is only able to capture non-trivial contributions from $\boldsymbol{T}^{(5)}$, while the PIML model, with the functional space enriched by a NN, is able to capture all the high-order contributions. Even though it seems that the average contributions (as measured by the Frobenius norm) are small for $\boldsymbol{T}^{(n)}, n \in 5,...10$, they are highly intermittent and may play an important role in improving the orientation of the eigenvectors, as observed in the Fig. \ref{fig:pdfangle1}.}

\begin{figure}[hbt]
	\includegraphics[width=5in]{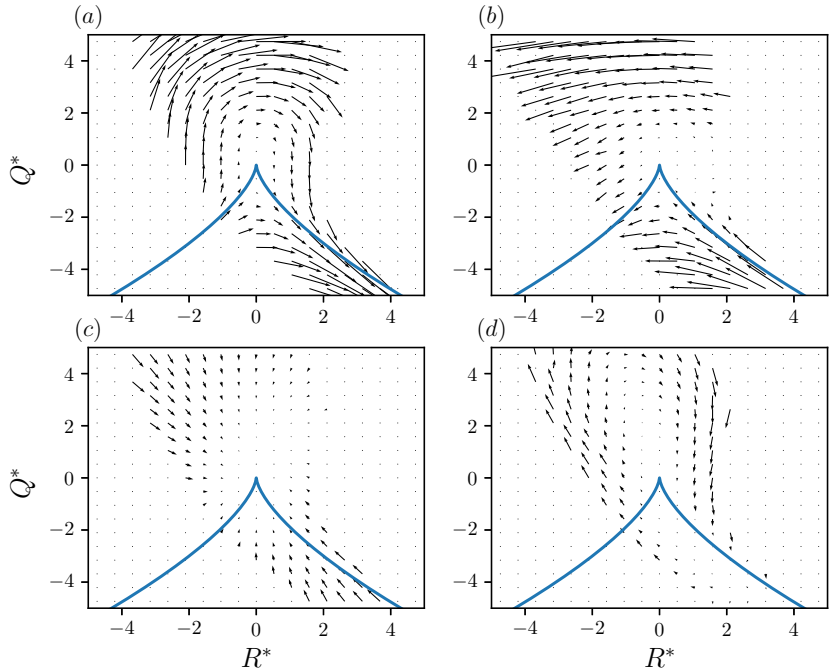}
	\caption{Lagrangian dynamics projected onto $Q-R$ phase plane: (\textit{a}) Restricted Euler term, (\textit{b}) pressure Hessian term, (\textit{c}) viscous term, and (\textit{d}) sum of all the terms. Zero discriminant line, $D=27 R^2-4 Q^3=0$, descriptive of the RE dynamics \citep{vieillefosse1982local}, is shown in blue.}
	\label{fig:qrdynam}
\end{figure}

This analysis confirms that by expanding the functional space using TBNN, the PIML model is capable of predicting rather accurately orientation/eigenvectors and magnitude/eigenvalues of the non-local pressure Hessian. Encouraged by these results, let us take a step further and study how the PIML model represents statistics of the VGT second invariant, $Q$, defined according to Eq.~(\ref{eqn:q}) and the third invariant, $R= -\frac{1}{3}A_{mn}A_{nl}A_{lm}$.  Specifically, we follow the $Q-R$ plane dynamics of the VGT,  widely used to describe the geometry of turbulent flows  \citep{chong1998turbulence,martin1998dynamics,ooi1999study,chertkov1999lagrangian,wang2012flow,chu2013topological,bechlars2017evolution,tian2019density}.
The Lagrangian dynamics, as described by Eq.~(\ref{eqn:vgtdynam}), can then be projected onto the $Q-R$ plane:
\begin{subequations}
\label{eqn:qrdynam}
\begin{eqnarray}
\frac{dQ}{dt}&=& -3R + A_{mn}H_{nm} + A_{mn}T_{nm} \\
\frac{dR}{dt}&=&\frac{2}{3}Q^2 +A_{mn}A_{nl}H_{lm} + A_{mn}A_{nl}T_{lm}
 \end{eqnarray}
\end{subequations}
Dynamics (rate of change) within the $(Q,R)$ plane is described, naturally, by the vector $(\frac{dQ}{dt},\frac{dR}{dt})$, formed from the components on the left hand sides of the equations. 
One can also form respective two-component vectors describing the $(Q, R)$ plane dynamics associated with the RE-, pressure Hessian-, and viscous- contributions on the right-hand sides of Eqs.~(\ref{eqn:qrdynam}). 
Fig. \ref{fig:qrdynam} shows the Conditional Mean Vector (CMV) plots of the projected dynamics.

Let us make a quick detour and comment on the dynamics of different vectors in the QR plane, as extracted from the DNS and shown in Fig.~\ref{fig:qrdynam}.  The Restricted Euler dynamics is deterministic
and, if no other terms are present, should move the "QR-particle" largely along the zero discriminant line. The RE term extracted from DNS, seen in Fig.~\ref{fig:qrdynam}(a), shows this as a general trend. As shown in Fig.~\ref{fig:qrdynam}(b), the pressure Hessian term counters the effects of Restricted Euler in some areas and moves the "QR-particle" towards the left part of the QR plane. The QR-vector associated with the viscous contribution largely points towards the origin, consistent with the fact that it aims to damp the turbulent fluctuations. 
The contribution summing up all the three vectors based on the right-hand side of Eqs.~(\ref{eqn:qrdynam}) and shown in Fig.~\ref{fig:qrdynam}(d) results in the classical clock-wise motion around the zero discriminant line resulting in the tear-drop shape of the QR plain PDF \citep{chertkov1999lagrangian}.  

\begin{figure}[hbt]
	\includegraphics[width=4in]{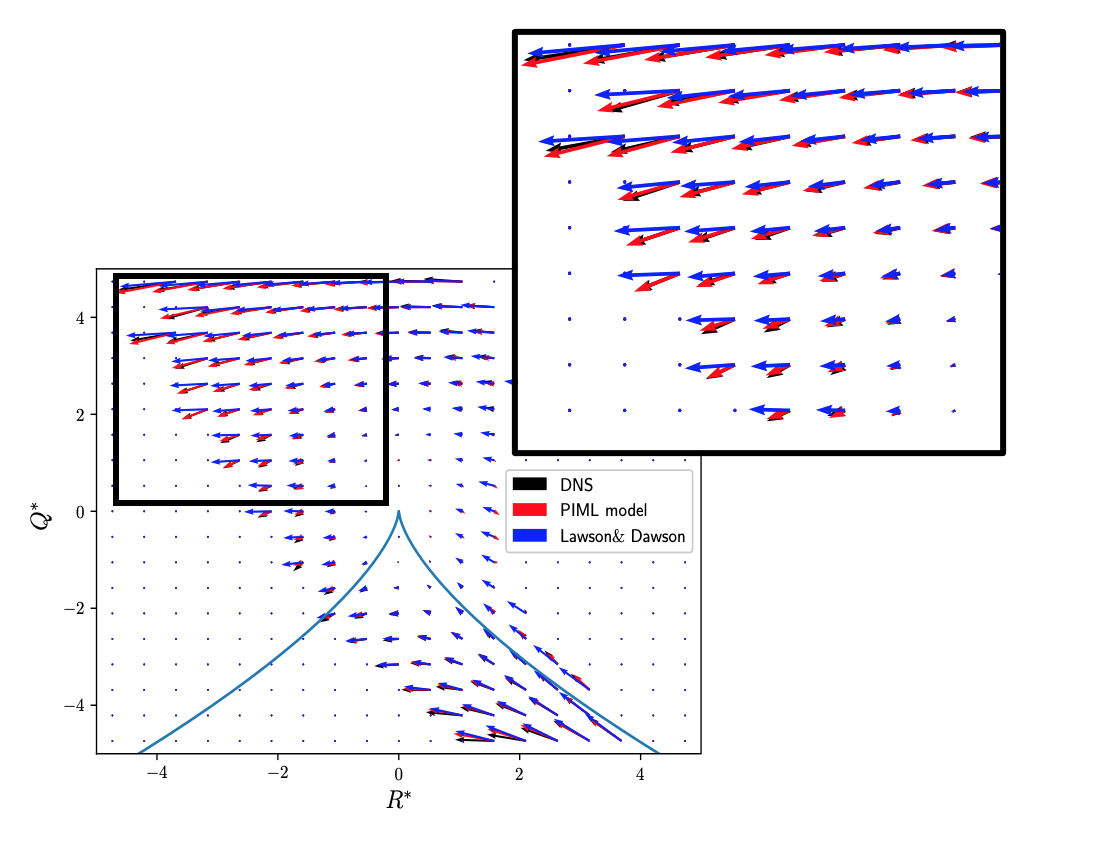}
	\caption{Comparison of pressure Hessian contributions on the $Q-R$ phase plane.}
	\label{fig:phesscompare}
\end{figure}

\begin{figure}[hbt]
	\includegraphics[width=5in]{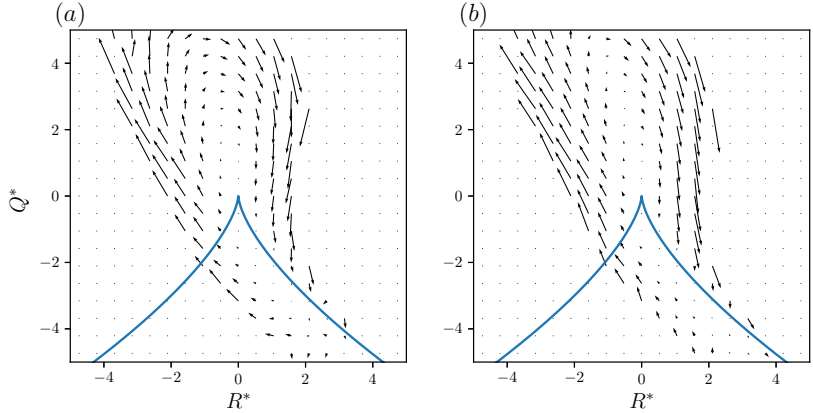}
	\caption{Comparison of Lagrangian dynamics between (\textit{a}) DNS and (\textit{b}) PIML model on the $Q-R$ phase plane.}
	\label{fig:qrdynamcompare}
\end{figure}

Returning back to the analysis of the PIML model performance, we now turn to the QR-plane comparison of PIML and also non-ML-based empirical model from \citet{lawson2015velocity} with DNS, shown in Fig.~\ref{fig:phesscompare}. Evidently, the PIML model improves over the non-ML-based model of \citet{lawson2015velocity} both in terms of representing magnitude and direction of the CMV, especially in the bottom left and bottom right quadrants. Substituting predictions of the PIML model for the right hand side terms in  Eq.~(\ref{eqn:qrdynam}) one can extract PIML-dynamics of the $(Q, R)$ vector shown in Fig.~\ref{fig:qrdynamcompare}. We observe that the PIML model results in a reasonably accurate reproduction of the clock-wise dynamics in the Q-R plane. 

Summarizing results discussed in this section -- we observe that the PIML model performs better than the reduced empirical models constructed so far, in terms of  reproducing reasonably well DNS results for Kolmogorov scale dynamics of the VGT tensor, expressed via magnitude and orientation of its eigenvalues and eigenvectors, and dynamics of its second and third invariants.


\subsection{{\em A Posteriori} Analysis}


After the {\em a priori} analysis of the PIML model, which has revealed its success in representing the DNS results, it seems natural for the next move to perform a more robust test consisting of substituting the PIML model for various terms on the right-hand side of the Lagrangian Eqs.~(\ref{eqn:vgtdynamics}). However, the non-local and stochastic nature of the contributions suggest that we should also augment the deterministic PIML terms with stochastic corrections.
This suggests we need to model the original unclosed Lagrangian ODE (Eq. \ref{eqn:vgtdynamics}) by a closed Stochastic ODE (SODE). Inspired by previous (however non-ML based) papers advancing this line of thought \citep{chertkov1999lagrangian,wilczek2014pressure,johnson2016closure} we consider the following stochastic model:
\begin{eqnarray}
   \frac{d A_{ij}}{dt} &=& E_{ij}+H_{ij,PIML}+T_{ij,model} + b_{ijmn}dW_{mn}, \label{eqn:sdevgtdynam}
\end{eqnarray}
where the newly introduced $\boldsymbol{b}$ is the covariance matrix of the stochastic (Wiener/Langevin) contribution to the dynamics. We borrow the form of the covariance matrix from \citet{johnson2016closure}, $b_{ijmn}=-\frac{1}{3}\sqrt{\frac{D_s}{5}}\delta_{ij}\delta_{mn}+\frac{1}{2}\left(\sqrt{\frac{D_s}{5}}+\sqrt{\frac{D_a}{3}}\right)\delta_{im}\delta_{jn} +\frac{1}{2}\left(\sqrt{\frac{D_s}{5}}-\sqrt{\frac{D_a}{3}}\right)\delta_{in}\delta_{jm}$ and learn the tuning parameters $D_a$ and $D_s$ from the residual between "ground-truth" and model prediction. \revision{In addition, we use a simple linear diffusion to model the viscous contributions $\boldsymbol{T}$, where the coefficients, defined as linear scalar function of invariants, are learned from the data. }

\begin{figure}[hbt]
	\includegraphics[width=6in]{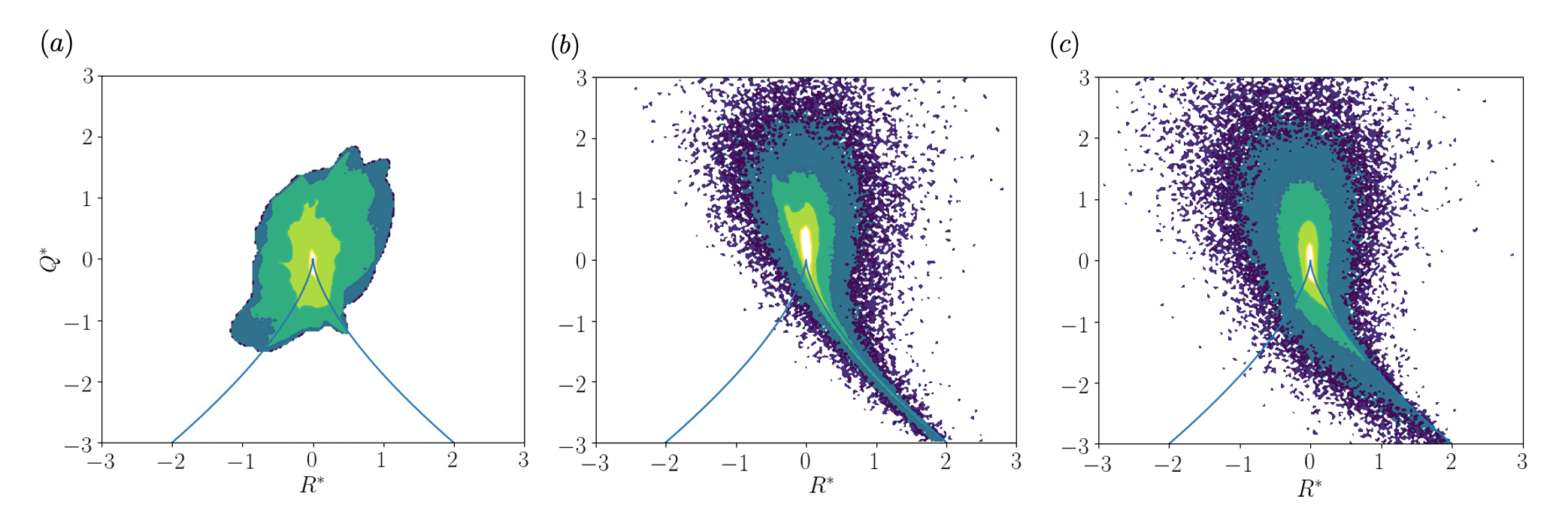}
	\caption{$Q-R$ plane PDF. 
	(\textit{a}) Random initialization (Gaussian, traceless VGT), (\textit{b}) PIML SODE prediction, and (\textit{c}) DNS. See text for details.}
	\label{fig:qrpdf}
\end{figure}

To have a preliminary assessment of the stochastic model validity we perform {\em a posteriori} test, consisting of seeding a large number (one million) of particles initialized with a random Gaussian VGT and then evolving each particle according to the SODE Eqs.~(\ref{eqn:sdevgtdynam}). In this experiment, we use a second-order predictor-corrector method to advance the particles in time.
The total integration is performed for half eddy turnover time, where the eddy turnover time is calculated using the formula $l/u'_{rms}$, where $l$ is the integral length scale of the flow. Fig.~\ref{fig:qrpdf}(b) shows the $(Q, R)$-plane of the PDF which we get for this experiment, as compared with the "ground truth" data shown in Fig.~\ref{fig:qrpdf}(c) and the result we get assuming a Gaussian i.i.d statistics for the components of VGT. Obviously, the Gaussian example shown in Fig.~\ref{fig:qrpdf}(a) (which we use to initiate the SODE PIML dynamics) is completely off in terms of the shape and spread of the joint distribution. However, comparison of the resulting SODE-map of the Gaussian seed, shown in Fig.~\ref{fig:qrpdf}(b),  with the "ground truth" data is satisfactory. The two share the similar tear-drop shape attributed to certain preferred local configurations associated with vortex stretching, and similar increased spread related to the large intermittency of small-scale turbulence.

\begin{figure}[hbt]
	\includegraphics[width=6in,height=2.5in]{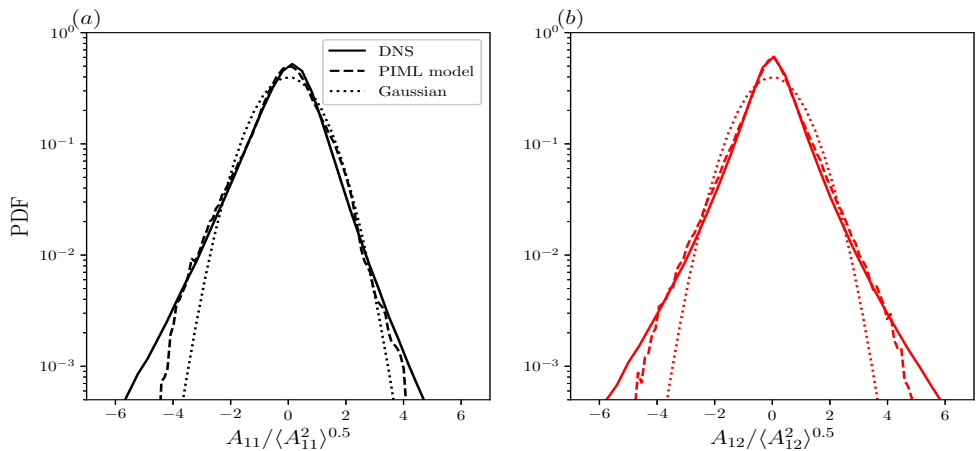}
	\caption{\revision{Standardized PDFs of the longitudinal component (a): $A_{11}$ and transverse component (b): $A_{12}$ of the VGT.}}
	\label{fig:pdfvgt}
\end{figure}

\revision{To further quantify the statistical properties of the VGT generated from the {\em a posteriori} test, we compare the standardized PDFs of the longitudinal, $A_{11}$, and transverse, $A_{12}$, components directly with those calculated from the DNS data, as well as their random Gaussian initialization. Evidently, our PIML SODE is capable, given the initial Gaussian field, of reproducing non-Gaussian features of the VGT. In particular, the skewed PDF of the longitudinal component can be captured, even though the skewness factor is under-predicted  (-0.3 for the PIML model and -0.55 for DNS). For the transverse component, the predicted PDF is symmetric (skewness factor around 0), which is in full agreement with the DNS data.}

\begin{figure}[hbt]
	\includegraphics[width=3in]{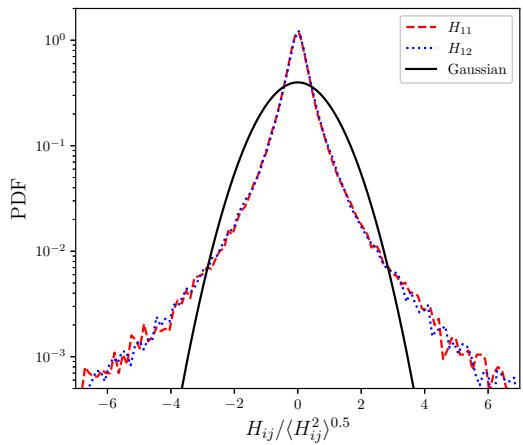}
	\caption{Standardized PDFs of the PIML model error of the longitudinal $H_{11}$ and transverse $H_{12}$ components of pressure Hessian.}
	\label{fig:errorpdf}
\end{figure}

\begin{figure}[hbt]
	\includegraphics[width=4in]{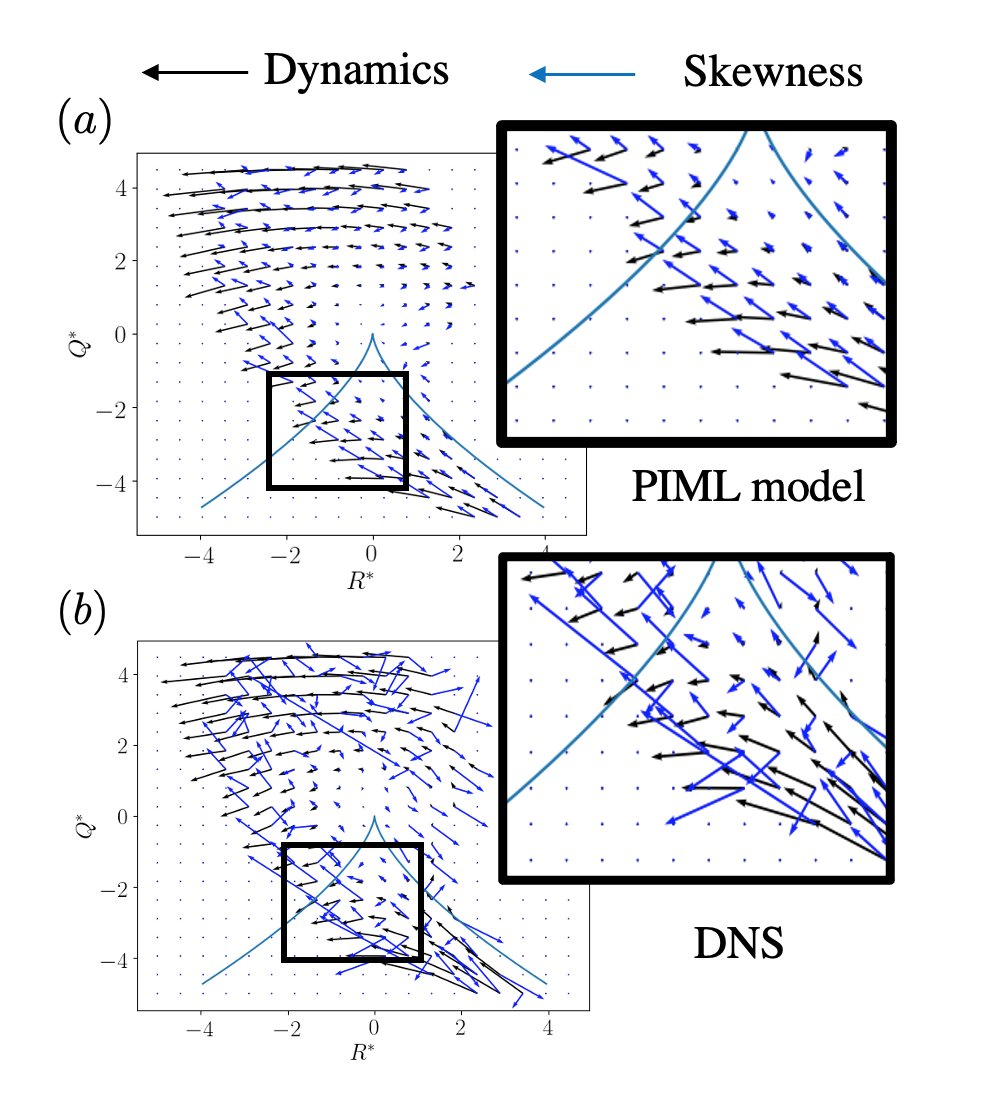}
	\caption{CMV of the PIML model prediction on the $Q-R$ phase plane. Both the mean (black arrows) and skewness (blue arrows) are shown with zoomed in view of the third and fourth quadrants.}
	\label{fig:cmvskewness}
\end{figure}

Upon further (more detailed) examination of the PIML results, we note that the PIML SODE predicts a narrower tail along the zero discriminant line compared to the DNS. We also observe that in the PIML-based experiments there are fewer particles ending in the left-bottom and right-bottom quadrants of the $Q-R$ plane.
As shown in Fig.~\ref{fig:qrdynam}, the contribution from the pressure Hessian plays a major role in moving QR-particles into the left-bottom and left-up quadrant. Besides,  it is known from  \citet{tian2019density} that the pressure Hessian has the largest magnitude, as compared with other contributions to the dynamics,  and it is also highly non-Gaussian (intermittent).  
These observations suggest that even though the stochastic PIML model is capable of predicting rather accurately the mean-value of the non-local pressure Hessian, it fails to represent correctly its higher-order moments, e.g. skewness, kurtosis, and tails of the pressure Hessian distribution. 
To make this statement clear, we show standardized error in the SODE-PIML prediction in Fig.~\ref{fig:errorpdf}. The error is calculated using the residual between pressure Hessian predicted by PIML model and "ground-truth" DNS, $\boldsymbol{H}_{DNS}-\boldsymbol{H}_{PIML}$. As expected,  the PDF shows tails that are much wider than one would expect from non-intermittent (Gaussian) statistics. We conclude that SODE-PIML misses rare events pushing more QR-particles into the left quadrants of the QR plane.
Misrepresentation of statistics by SODE-PIML is further quantified in Fig.~\ref{fig:cmvskewness}, where we show in addition to the mean vector (black), the respective vector associated with the skewness (blue) of the QR-plane vector associated with the right-hand side of Eqs.~(\ref{eqn:vgtdynamics2}). We clearly see here a rather significant mismatch in skewness between the SODE-PIML and the DNS results observed in the left part of the $Q-R$ plane. 

Several factors could contribute to the fact that the PIML model fails to predict these high-order non-Gaussian statistics: the choice of the loss function and the choice of stochastic terms. By using the quadratic loss function and a Gaussian stochastic term, the error is essentially assumed to be Gaussian and therefore high-order statistics cannot be represented. Future work on developing better loss functions and stochastic term will be discussed in the last section. \revision{Another caveat of the constructed SODE is that the viscous contributions are approximated using a simple linear diffusion model, which is generally not enough to damp high-order nonlinear contributions from the ML models. This is a common issue of all the ML models. It is generally difficult to devise a high-order nonlinear damping term, like the one proposed in \citep{leppin2020capturing}, to prevent instability of the  Lagrangian particle evolution. We are working on improving stability of the Lagrrangian dynamics taking advantage of a number of  recently developed ML techniques.} 

\section{{\em A Priori} Analysis of Coarse-Grained Velocity Gradient Tensor \label{sec:cvgt}}


We will start the discussion of the process of coarse-graining with a disclaimer. Coarse-graining/filtering of the Lagrangian dynamics of VGT  adds a "geometrical" complexity to the problem, in the sense that even if it is spatially isotropic,  i.e. filtering is over a ball of a fixed size, the ball will evolve acquiring a complicated, typically multi-scale shape. This is the foundation for some of the advanced physics-based Lagrangian models \citep{chertkov1999lagrangian,johnson2016closure}. Having acknowledged the complexity of this effect associated with the Lagrangian description, we will not address it here in full. Instead, we limit the description to only spatial filtering over a static and isotropic volume and focus instead on extending modeling of the small-scale VGT based on locality to that of the coarse-grained VGT on a fixed volume. 


\begin{figure}[hbt]
	\includegraphics[width=5in]{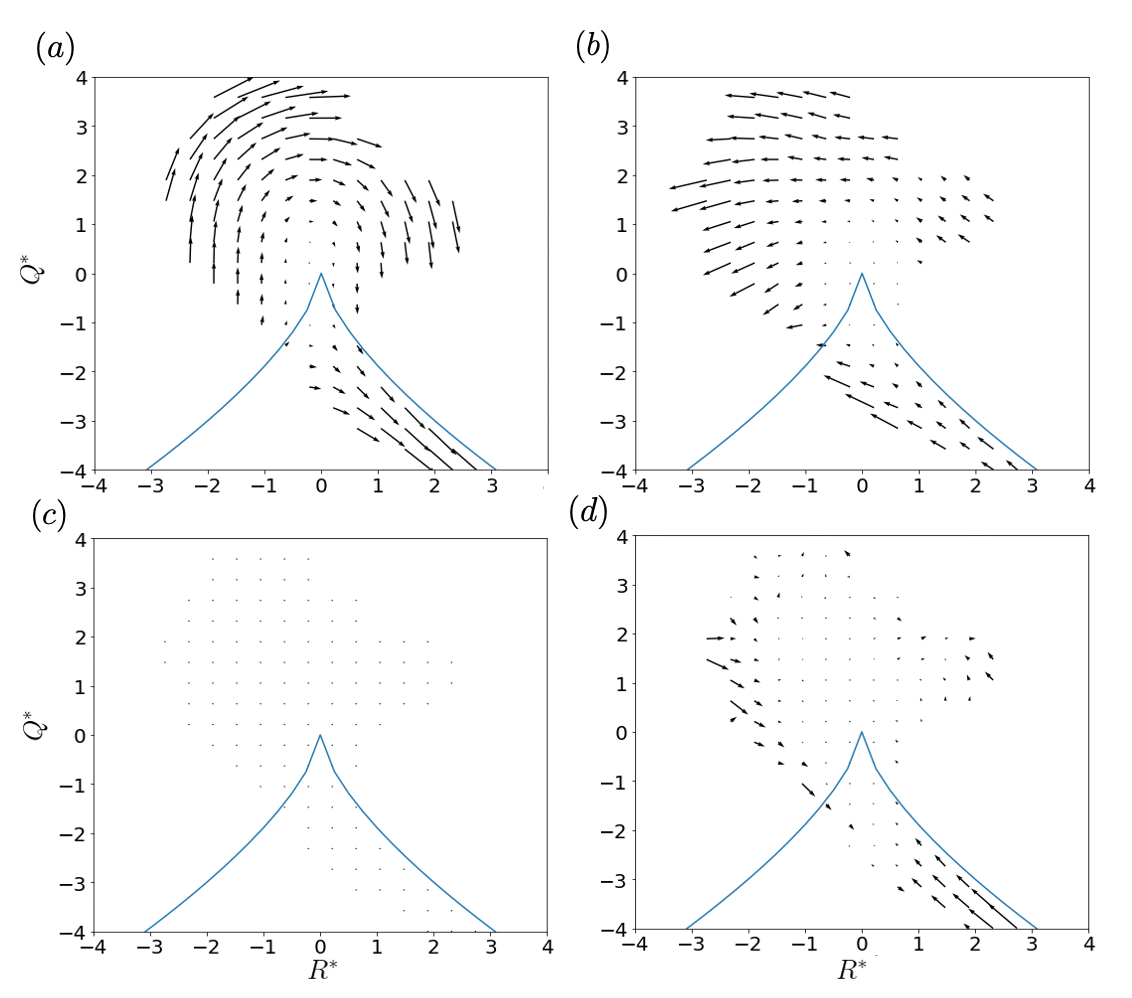}
	\caption{CMV of the coarse-grained VGT dynamics with filter scale selected to be, $l_{\Delta} = \pi/8$. (\textit{a}) Restricted Euler, (\textit{b}) pressure Hessian, (\textit{c}) viscous term, and (\textit{d}) sub-filter contribution.}
	\label{fig:cmvcoarsevgt}
\end{figure}

We start this discussion with an analysis, illustrated in Fig.~\ref{fig:cmvcoarsevgt}, of what DNS shows for different terms on the right-hand side of Eqs.~(\ref{eqn:vgtdynamics3}) where the scale of coarse-graining is selected to be $l_{\Delta} = \pi/8$, i.e. somewhere within the inertial interval limited by the energy-containing scale, estimated by $\pi$,  and the viscous/Kolmogorov scale, estimated by $\pi/512$. Note that the velocity forcing used here, following \citet{petersen2010forcing,daniel2018reaction}, enforces the Kolmogorov scale at the onset, while energy is added at wavenumbers $k\leq |1.5|$.
We observe that the QR-flow of the coarse-grained RE term and the coarse-grained pressure Hessian term are similar to the respective unfiltered terms. We also observe that the viscous contribution becomes small and its "eddy dissipation" role is taken by the sub-filter term. 

\begin{figure}[hbt]
	\includegraphics[width=4in]{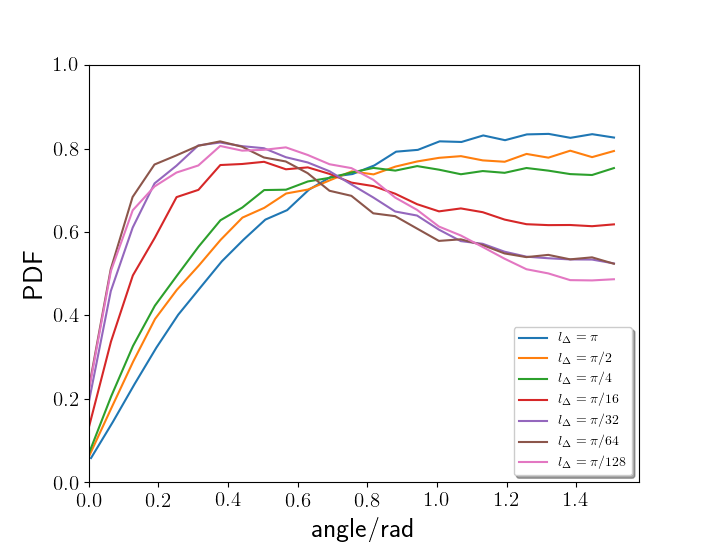}
	\caption{PDF of the angle (radian) between the predicted $e_1$ eigenvector and DNS data for various filter sizes.}
	\label{fig:pdfanglecoarse}
\end{figure}

Now we turn our attention to the focal point of this section -- empirical modeling with a NN of the pressure Hessian and sub-filter contributions. 
The coarse-grained pressure Hessian is modeled using the same TBNN with the symmetric integrity bases as used above in the unfiltered (single-point) case. We consider coarse-graining at different scales independently, therefore train the TBNN models for each scale anew. 
Fig.~\ref{fig:pdfanglecoarse} shows the alignment between model prediction and DNS for the pressure Hessian coarse-grained at different scales. We note that the predicted alignment is not as good as that in the unfiltered PIML model. This means that adding a lengthscale to the dynamics makes it harder for predicting the orientation. As expected, the model performance converges towards the unfiltered model as the filter size decreases. Table \ref{tab:eigvalcoarse} shows correlation coefficients between eigenvalues resulting from the PIML model and DNS data coarse-grained at different scales. We note that the performance of the same TBNN architecture becomes worse as the filter size increases. This analysis suggests that the TBNN architecture is less suitable for learning coarse-grained dynamics as the scale of coarse-graining departs significantly from the viscous scale. 

\begin{table*}
\caption{\label{tab:eigvalcoarse}%
Correlation coefficients of the eigenvalues for different scales of coarse-graining.
}
\begin{ruledtabular}
\begin{tabular}{cccc}
\textrm{Filter size}&
\textrm{corr($e_{1,model},e_{1,DNS}$)}&
\textrm{corr($e_{2,model},e_{2,DNS}$)} & 
\textrm{corr($e_{3,model},e_{3,DNS}$)} \\
\colrule
$l_\Delta = \pi$ & 0.17 & 0.21 & 0.31\\
$l_\Delta = \pi/2$ & 0.35 & 0.07 & 0.33\\
$l_\Delta = \pi/4$ & 0.50 & 0.008 & 0.51\\
$l_\Delta = \pi/16$ & 0.61 & 0.06 & 0.63\\
$l_\Delta = \pi/32$ & 0.66 & 0.12 & 0.67\\
$l_\Delta = \pi/64$ & 0.71 & 0.16 & 0.72\\
$l_\Delta = \pi/128$ & 0.72 & 0.20 & 0.73\\

\end{tabular}
\end{ruledtabular}
\end{table*}

\begin{figure}[hbt]
	\includegraphics[width=3.5in]{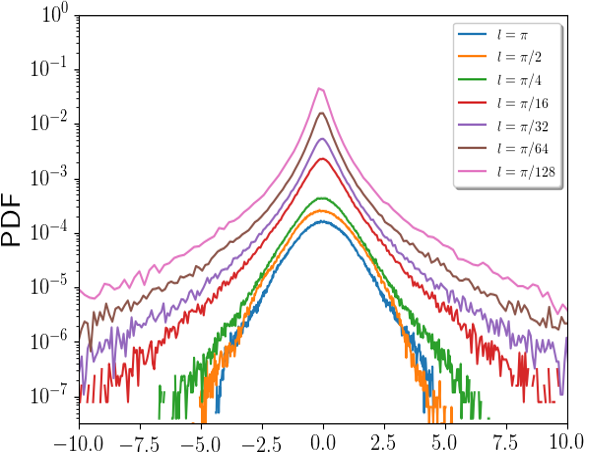}
	\caption{PDF of the error of PIML model prediction for various filter sizes. The curves are re-normalized to avoid clustering.}
	\label{fig:errorpdfcoarse}
\end{figure}

Fig.~\ref{fig:errorpdfcoarse} shows PDFs of prediction error for PIML models of different coarse-grained scales. Interestingly, the error becomes more and more Gaussian-like when the filter size increases. This indicates that the Langevin type SODE model is better suited for modeling coarse-grained dynamics than those occurring at small-scales.

\begin{figure}[hbt]
	\includegraphics[width=4in]{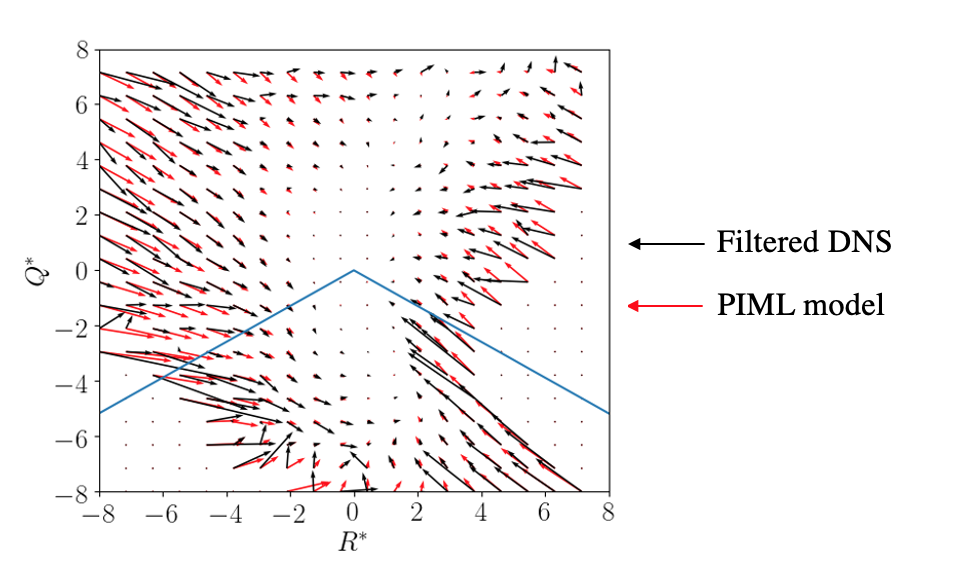}
	\caption{Comparison of sub-filter CMV between the PIML model prediction and DNS. PIML model prediction is shown as red and DNS data is shown as black. The chosen coarse-graining size is $\pi/8$.}
	\label{fig:cmvsubfiltercompare}
\end{figure}

The sub-filter contribution is modeled using the extended TBNN framework with all 16 integrity bases (10 symmetric and 6 skew symmetric) included. Fig. \ref{fig:cmvsubfiltercompare} compares the CMV of sub-filter contribution for coarse-graining size $l_\Delta=\pi/8$. The PIML model can generally predict the qualitative behavior of the dynamics, but there still exists a large error in the magnitude, especially in the third quadrant. On top of that, the lack of explicit length scale information in the model makes it harder to generalize. This indicates an improved modeling strategy is required for coarse-grained dynamics.

\section{Conclusions and Future Work}
\label{sec:conclusion}

Modeling Velocity Gradient Tensor (VGT), evaluated locally at a point in the Eulerian and/or Lagrangian frames, and/or averaged over scales within the inertial range of turbulence, is the focus of this manuscript. We are not the first to study this interesting object, which is known to be important for understanding quantitatively and qualitatively many principal features of developed incompressible turbulence,  such as geometry and statistics of energy transfer,  vorticity-strain alignment, mixing, etc. In this work, we aim at developing a reduced empirical model of a stochastic ODE type describing the evolution of VGT at both small-scales and coarse-grained scales. We take advantage of the emerging Machine Learning techniques/tools and inject them with physics-inspired and mathematics-based constraints to develop a mixed approach - Physics Informed Machine Learning. This allowed us to go much deeper than before in terms of quality achieved in reproducing VGT with a reduced model. We show in this paper that the PIML approach applied to the Lagrangian evolution of the coarse-grained VGT is advantageous over existing methods. To achieve this conclusion, it required overcoming a number of obstacles which we find useful to list and re-discuss below.


A major difficulty in modeling dynamics of the VGT in incompressible flow is related to its non-locality of pressure contributions inherited from the principal non-linearity of the microscopic description. We tackle this problem by proposing an empirical model for the spatial distribution of the second invariant of VGT that satisfies Galilean invariance, rotational invariance, and incompressibility condition. After utilizing an ansatz for non-local pressure Hessian, the PIML model is re-formatted into a combination of functions of invariants and integrity bases. This functional form falls naturally into the TBNN architecture with physically interpretable parameterization. After training the model with DNS data, we arrived at an improved representation of the magnitude and orientation of pressure Hessian eigenvales and eigenvectors. In contrast to Physics-Blind ML model, we can interpret such improvement by analyzing the statistical properties of scalar coefficients for different tensor bases. A stochastic ODE is constructed to model the full evolution of VGT with a simplified constant co-variance matrix for the stochastic term. An {\em a posteriori} test is then conducted by temporally evolving the SODE from a randomly Gaussian initial field. The classical tear-drop shape of the $Q-R$ joint distribution is reproduced, but the tail is more elongated compared to the DNS data. Further diagnostics show that the non-Gaussian statistics of the pressure Hessian cannot be fully represented by the PIML model. 

For the coarse-grained VGT dynamics, a similar approach is adopted using the extended TBNN framework, which expands the integrity bases for skew-symmetric tensor prediction. The PIML model is trained using filtered DNS data with different filter sizes. With increasing filter size, the dynamics become more Gaussian-like, which makes stochastic modeling more reasonable. However, the performance of the model, i.e. orientation and magnitude of the eigenvectors and eigenvalues of the coarse-grained VGT, is worse compared to the unfiltered case. This is likely because the current extended TBNN is not able to embed length-scale information in the model.

We also acknowledge that our current PIML model simplifies the spatial complexity due to the single-point modeling strategy. We recognize that better spatial modeling should be Lagrangian, injecting into consideration the dynamics of the coarse-grained volume. 

To correct the problems with the VGT-only reduced modeling encountered in this paper, we have started to work in \citep{criston2020} on extending it.  In particular, and acting consistently with the tetrad model approach developed in \citet{chertkov1999lagrangian}, we work on including into Lagrangian description geometrical characteristics, such as the shape of the coarse-graining fluid blob co-evolving with the VGT. When training on high-Reynolds number DNS data, we expect the improved model to learn the universal inertial-range Lagrangian dynamics. The small-scale VGT dynamics with a high dependence on Reynolds number can then be modeled using the inter-scale interaction and energy transfer.
We also work on improving and tuning the ML approaches. Thus, we have developed a Bayesian learning approach towards constructing SODE by learning (and not postulating as above) the co-variance matrix (possibly dependent on both the coarse-graining shape and scale and on the coarse-grained VGT).  We also envision exploring in the future ways to provide a better representation for the high-order/non-Gaussian statistics of turbulence. Injecting non-Wiener/non-Langevin noise and working with a richer family of loss functions  (used to describe mismatch between DNS and empirical results) may help to tackle the problem of better representation of the turbulence and the inherited intermittency. 

\begin{acknowledgments}
This work was performed under the auspices of DOE. Financial support comes from Los Alamos National Laboratory (LANL), Laboratory Directed Research and Development (LDRD) project "MELT: Machine Learning for Turbulence," 20190059DR and its subcontract to UArizona. LANL, an affirmative action/equal opportunity employer, is managed by Triad National Security, LLC, for the National Nuclear Security Administration of the U.S. Department of Energy under contract 89233218CNA000001.
\end{acknowledgments}

\bibliography{main}

\begin{thebibliography}{53}%
\makeatletter
\providecommand \@ifxundefined [1]{%
 \@ifx{#1\undefined}
}%
\providecommand \@ifnum [1]{%
 \ifnum #1\expandafter \@firstoftwo
 \else \expandafter \@secondoftwo
 \fi
}%
\providecommand \@ifx [1]{%
 \ifx #1\expandafter \@firstoftwo
 \else \expandafter \@secondoftwo
 \fi
}%
\providecommand \natexlab [1]{#1}%
\providecommand \enquote  [1]{``#1''}%
\providecommand \bibnamefont  [1]{#1}%
\providecommand \bibfnamefont [1]{#1}%
\providecommand \citenamefont [1]{#1}%
\providecommand \href@noop [0]{\@secondoftwo}%
\providecommand \href [0]{\begingroup \@sanitize@url \@href}%
\providecommand \@href[1]{\@@startlink{#1}\@@href}%
\providecommand \@@href[1]{\endgroup#1\@@endlink}%
\providecommand \@sanitize@url [0]{\catcode `\\12\catcode `\$12\catcode
  `\&12\catcode `\#12\catcode `\^12\catcode `\_12\catcode `\%12\relax}%
\providecommand \@@startlink[1]{}%
\providecommand \@@endlink[0]{}%
\providecommand \url  [0]{\begingroup\@sanitize@url \@url }%
\providecommand \@url [1]{\endgroup\@href {#1}{\urlprefix }}%
\providecommand \urlprefix  [0]{URL }%
\providecommand \Eprint [0]{\href }%
\providecommand \doibase [0]{https://doi.org/}%
\providecommand \selectlanguage [0]{\@gobble}%
\providecommand \bibinfo  [0]{\@secondoftwo}%
\providecommand \bibfield  [0]{\@secondoftwo}%
\providecommand \translation [1]{[#1]}%
\providecommand \BibitemOpen [0]{}%
\providecommand \bibitemStop [0]{}%
\providecommand \bibitemNoStop [0]{.\EOS\space}%
\providecommand \EOS [0]{\spacefactor3000\relax}%
\providecommand \BibitemShut  [1]{\csname bibitem#1\endcsname}%
\let\auto@bib@innerbib\@empty
\bibitem [{\citenamefont {Ling}\ \emph {et~al.}(2016)\citenamefont {Ling},
  \citenamefont {Kurzawski},\ and\ \citenamefont
  {Templeton}}]{ling2016reynolds}%
  \BibitemOpen
  \bibfield  {author} {\bibinfo {author} {\bibfnamefont {J.}~\bibnamefont
  {Ling}}, \bibinfo {author} {\bibfnamefont {A.}~\bibnamefont {Kurzawski}},
  and\ \bibinfo {author} {\bibfnamefont {J.}~\bibnamefont {Templeton}},\
  }\bibfield  {title} {\bibinfo {title} {Reynolds averaged turbulence modelling
  using deep neural networks with embedded invariance},\ }\href@noop {}
  {\bibfield  {journal} {\bibinfo  {journal} {J. Fluid Mech.}\ }\textbf
  {\bibinfo {volume} {807}},\ \bibinfo {pages} {155} (\bibinfo {year}
  {2016})}\BibitemShut {NoStop}%
\bibitem [{\citenamefont {Meneveau}(2011)}]{meneveau2011lagrangian}%
  \BibitemOpen
  \bibfield  {author} {\bibinfo {author} {\bibfnamefont {C.}~\bibnamefont
  {Meneveau}},\ }\bibfield  {title} {\bibinfo {title} {Lagrangian dynamics and
  models of the velocity gradient tensor in turbulent flows},\ }\href@noop {}
  {\bibfield  {journal} {\bibinfo  {journal} {Annu. Rev. Fluid Mech.}\ }\textbf
  {\bibinfo {volume} {43}},\ \bibinfo {pages} {219} (\bibinfo {year}
  {2011})}\BibitemShut {NoStop}%
\bibitem [{\citenamefont {Eyink}(2006)}]{eyink2006multi}%
  \BibitemOpen
  \bibfield  {author} {\bibinfo {author} {\bibfnamefont {G.~L.}\ \bibnamefont
  {Eyink}},\ }\bibfield  {title} {\bibinfo {title} {Multi-scale gradient
  expansion of the turbulent stress tensor},\ }\href@noop {} {\bibfield
  {journal} {\bibinfo  {journal} {J. Fluid Mech.}\ }\textbf {\bibinfo {volume}
  {549}},\ \bibinfo {pages} {159} (\bibinfo {year} {2006})}\BibitemShut
  {NoStop}%
\bibitem [{\citenamefont {Johnson}\ and\ \citenamefont
  {Meneveau}(2017)}]{johnson2017turbulence}%
  \BibitemOpen
  \bibfield  {author} {\bibinfo {author} {\bibfnamefont {P.~L.}\ \bibnamefont
  {Johnson}}and\ \bibinfo {author} {\bibfnamefont {C.}~\bibnamefont
  {Meneveau}},\ }\bibfield  {title} {\bibinfo {title} {Turbulence intermittency
  in a multiple-time-scale navier-stokes-based reduced model},\ }\href@noop {}
  {\bibfield  {journal} {\bibinfo  {journal} {Phys. Rev. Fluids}\ }\textbf
  {\bibinfo {volume} {2}},\ \bibinfo {pages} {072601} (\bibinfo {year}
  {2017})}\BibitemShut {NoStop}%
\bibitem [{\citenamefont {Doan}\ \emph {et~al.}(2018)\citenamefont {Doan},
  \citenamefont {Swaminathan}, \citenamefont {Davidson},\ and\ \citenamefont
  {Tanahashi}}]{doan2018scale}%
  \BibitemOpen
  \bibfield  {author} {\bibinfo {author} {\bibfnamefont {N.~A.~K.}\
  \bibnamefont {Doan}}, \bibinfo {author} {\bibfnamefont {N.}~\bibnamefont
  {Swaminathan}}, \bibinfo {author} {\bibfnamefont {P.}~\bibnamefont
  {Davidson}}, and\ \bibinfo {author} {\bibfnamefont {M.}~\bibnamefont
  {Tanahashi}},\ }\bibfield  {title} {\bibinfo {title} {Scale locality of the
  energy cascade using real space quantities},\ }\href@noop {} {\bibfield
  {journal} {\bibinfo  {journal} {Phys. Rev. Fluids}\ }\textbf {\bibinfo
  {volume} {3}},\ \bibinfo {pages} {084601} (\bibinfo {year}
  {2018})}\BibitemShut {NoStop}%
\bibitem [{\citenamefont {Carbone}\ and\ \citenamefont
  {Bragg}(2020)}]{carbone2020vortex}%
  \BibitemOpen
  \bibfield  {author} {\bibinfo {author} {\bibfnamefont {M.}~\bibnamefont
  {Carbone}}and\ \bibinfo {author} {\bibfnamefont {A.~D.}\ \bibnamefont
  {Bragg}},\ }\bibfield  {title} {\bibinfo {title} {Is vortex stretching the
  main cause of the turbulent energy cascade?},\ }\href@noop {} {\bibfield
  {journal} {\bibinfo  {journal} {J. Fluid Mech.}\ }\textbf {\bibinfo {volume}
  {883}} (\bibinfo {year} {2020})}\BibitemShut {NoStop}%
\bibitem [{\citenamefont {Johnson}(2020)}]{johnson2020energy}%
  \BibitemOpen
  \bibfield  {author} {\bibinfo {author} {\bibfnamefont {P.~L.}\ \bibnamefont
  {Johnson}},\ }\bibfield  {title} {\bibinfo {title} {Energy transfer from
  large to small scales in turbulence by multiscale nonlinear strain and
  vorticity interactions},\ }\href@noop {} {\bibfield  {journal} {\bibinfo
  {journal} {Phys. Rev. Lett.}\ }\textbf {\bibinfo {volume} {124}},\ \bibinfo
  {pages} {104501} (\bibinfo {year} {2020})}\BibitemShut {NoStop}%
\bibitem [{\citenamefont {Perry}\ and\ \citenamefont
  {Chong}(1987)}]{perry1987description}%
  \BibitemOpen
  \bibfield  {author} {\bibinfo {author} {\bibfnamefont {A.~E.}\ \bibnamefont
  {Perry}}and\ \bibinfo {author} {\bibfnamefont {M.~S.}\ \bibnamefont
  {Chong}},\ }\bibfield  {title} {\bibinfo {title} {A description of eddying
  motions and flow patterns using critical-point concepts},\ }\href@noop {}
  {\bibfield  {journal} {\bibinfo  {journal} {Annu. Rev. Fluid Mech.}\ }\textbf
  {\bibinfo {volume} {19}},\ \bibinfo {pages} {125} (\bibinfo {year}
  {1987})}\BibitemShut {NoStop}%
\bibitem [{\citenamefont {Chong}\ \emph {et~al.}(1990)\citenamefont {Chong},
  \citenamefont {Perry},\ and\ \citenamefont {Cantwell}}]{chong1990general}%
  \BibitemOpen
  \bibfield  {author} {\bibinfo {author} {\bibfnamefont {M.~S.}\ \bibnamefont
  {Chong}}, \bibinfo {author} {\bibfnamefont {A.~E.}\ \bibnamefont {Perry}},
  and\ \bibinfo {author} {\bibfnamefont {B.~J.}\ \bibnamefont {Cantwell}},\
  }\bibfield  {title} {\bibinfo {title} {A general classification of
  three-dimensional flow fields},\ }\href@noop {} {\bibfield  {journal}
  {\bibinfo  {journal} {Phys.~Fluids A: Fluid Dynam.}\ }\textbf {\bibinfo
  {volume} {2}},\ \bibinfo {pages} {765} (\bibinfo {year} {1990})}\BibitemShut
  {NoStop}%
\bibitem [{\citenamefont {Chong}\ \emph {et~al.}(1998)\citenamefont {Chong},
  \citenamefont {Soria}, \citenamefont {Perry}, \citenamefont {Chacin},
  \citenamefont {Cantwell},\ and\ \citenamefont {Na}}]{chong1998turbulence}%
  \BibitemOpen
  \bibfield  {author} {\bibinfo {author} {\bibfnamefont {M.~S.}\ \bibnamefont
  {Chong}}, \bibinfo {author} {\bibfnamefont {J.}~\bibnamefont {Soria}},
  \bibinfo {author} {\bibfnamefont {A.~E.}\ \bibnamefont {Perry}}, \bibinfo
  {author} {\bibfnamefont {J.}~\bibnamefont {Chacin}}, \bibinfo {author}
  {\bibfnamefont {B.~J.}\ \bibnamefont {Cantwell}}, and\ \bibinfo {author}
  {\bibfnamefont {Y.}~\bibnamefont {Na}},\ }\bibfield  {title} {\bibinfo
  {title} {Turbulence structures of wall-bounded shear flows found using dns
  data},\ }\href@noop {} {\bibfield  {journal} {\bibinfo  {journal} {J.~Fluid
  Mech.}\ }\textbf {\bibinfo {volume} {357}},\ \bibinfo {pages} {225} (\bibinfo
  {year} {1998})}\BibitemShut {NoStop}%
\bibitem [{\citenamefont {Martin}\ \emph
  {et~al.}(1998{\natexlab{a}})\citenamefont {Martin}, \citenamefont {Ooi},
  \citenamefont {Chong},\ and\ \citenamefont {Soria}}]{martin1998dynamics}%
  \BibitemOpen
  \bibfield  {author} {\bibinfo {author} {\bibfnamefont {J.}~\bibnamefont
  {Martin}}, \bibinfo {author} {\bibfnamefont {A.}~\bibnamefont {Ooi}},
  \bibinfo {author} {\bibfnamefont {M.~S.}\ \bibnamefont {Chong}}, and\
  \bibinfo {author} {\bibfnamefont {J.}~\bibnamefont {Soria}},\ }\bibfield
  {title} {\bibinfo {title} {Dynamics of the velocity gradient tensor
  invariants in isotropic turbulence},\ }\href@noop {} {\bibfield  {journal}
  {\bibinfo  {journal} {Phys. Fluids}\ }\textbf {\bibinfo {volume} {10}},\
  \bibinfo {pages} {2336} (\bibinfo {year} {1998}{\natexlab{a}})}\BibitemShut
  {NoStop}%
\bibitem [{\citenamefont {Ooi}\ \emph {et~al.}(1999)\citenamefont {Ooi},
  \citenamefont {Martin}, \citenamefont {Soria},\ and\ \citenamefont
  {Chong}}]{ooi1999study}%
  \BibitemOpen
  \bibfield  {author} {\bibinfo {author} {\bibfnamefont {A.}~\bibnamefont
  {Ooi}}, \bibinfo {author} {\bibfnamefont {J.}~\bibnamefont {Martin}},
  \bibinfo {author} {\bibfnamefont {J.}~\bibnamefont {Soria}}, and\ \bibinfo
  {author} {\bibfnamefont {M.~S.}\ \bibnamefont {Chong}},\ }\bibfield  {title}
  {\bibinfo {title} {A study of the evolution and characteristics of the
  invariants of the velocity-gradient tensor in isotropic turbulence},\
  }\href@noop {} {\bibfield  {journal} {\bibinfo  {journal} {J.~Fluid Mech.}\
  }\textbf {\bibinfo {volume} {381}},\ \bibinfo {pages} {141} (\bibinfo {year}
  {1999})}\BibitemShut {NoStop}%
\bibitem [{\citenamefont {Wang}\ and\ \citenamefont {Lu}(2012)}]{wang2012flow}%
  \BibitemOpen
  \bibfield  {author} {\bibinfo {author} {\bibfnamefont {L.}~\bibnamefont
  {Wang}}and\ \bibinfo {author} {\bibfnamefont {X.~Y.}\ \bibnamefont {Lu}},\
  }\bibfield  {title} {\bibinfo {title} {Flow topology in compressible
  turbulent boundary layer},\ }\href@noop {} {\bibfield  {journal} {\bibinfo
  {journal} {J.~Fluid Mech.}\ }\textbf {\bibinfo {volume} {703}},\ \bibinfo
  {pages} {255} (\bibinfo {year} {2012})}\BibitemShut {NoStop}%
\bibitem [{\citenamefont {Chu}\ and\ \citenamefont
  {Lu}(2013)}]{chu2013topological}%
  \BibitemOpen
  \bibfield  {author} {\bibinfo {author} {\bibfnamefont {Y.~B.}\ \bibnamefont
  {Chu}}and\ \bibinfo {author} {\bibfnamefont {X.~Y.}\ \bibnamefont {Lu}},\
  }\bibfield  {title} {\bibinfo {title} {Topological evolution in compressible
  turbulent boundary layers},\ }\href@noop {} {\bibfield  {journal} {\bibinfo
  {journal} {J.~Fluid Mech.}\ }\textbf {\bibinfo {volume} {733}},\ \bibinfo
  {pages} {414} (\bibinfo {year} {2013})}\BibitemShut {NoStop}%
\bibitem [{\citenamefont {Bechlars}\ and\ \citenamefont
  {Sandberg}(2017)}]{bechlars2017evolution}%
  \BibitemOpen
  \bibfield  {author} {\bibinfo {author} {\bibfnamefont {P.}~\bibnamefont
  {Bechlars}}and\ \bibinfo {author} {\bibfnamefont {R.~D.}\ \bibnamefont
  {Sandberg}},\ }\bibfield  {title} {\bibinfo {title} {Evolution of the
  velocity gradient tensor invariant dynamics in a turbulent boundary layer},\
  }\href@noop {} {\bibfield  {journal} {\bibinfo  {journal} {J.~Fluid Mech.}\
  }\textbf {\bibinfo {volume} {815}},\ \bibinfo {pages} {223} (\bibinfo {year}
  {2017})}\BibitemShut {NoStop}%
\bibitem [{\citenamefont {Tian}\ \emph {et~al.}(2019)\citenamefont {Tian},
  \citenamefont {Jaberi},\ and\ \citenamefont {Livescu}}]{tian2019density}%
  \BibitemOpen
  \bibfield  {author} {\bibinfo {author} {\bibfnamefont {Y.}~\bibnamefont
  {Tian}}, \bibinfo {author} {\bibfnamefont {F.~A.}\ \bibnamefont {Jaberi}},
  and\ \bibinfo {author} {\bibfnamefont {D.}~\bibnamefont {Livescu}},\
  }\bibfield  {title} {\bibinfo {title} {Density effects on the post-shock
  turbulence structure and dynamics},\ }\href@noop {} {\bibfield  {journal}
  {\bibinfo  {journal} {J. Fluid Mech.}\ }\textbf {\bibinfo {volume} {880}},\
  \bibinfo {pages} {935} (\bibinfo {year} {2019})}\BibitemShut {NoStop}%
\bibitem [{\citenamefont {Das}\ and\ \citenamefont
  {Girimaji}(2020)}]{das2020characterization}%
  \BibitemOpen
  \bibfield  {author} {\bibinfo {author} {\bibfnamefont {R.}~\bibnamefont
  {Das}}and\ \bibinfo {author} {\bibfnamefont {S.~S.}\ \bibnamefont
  {Girimaji}},\ }\bibfield  {title} {\bibinfo {title} {Characterization of
  velocity-gradient dynamics in incompressible turbulence using local
  streamline geometry},\ }\href@noop {} {\bibfield  {journal} {\bibinfo
  {journal} {arXiv preprint arXiv:2002.00259}\ } (\bibinfo {year}
  {2020})}\BibitemShut {NoStop}%
\bibitem [{\citenamefont {Tom}\ \emph {et~al.}(2021)\citenamefont {Tom},
  \citenamefont {Carbone},\ and\ \citenamefont {Bragg}}]{tom2021exploring}%
  \BibitemOpen
  \bibfield  {author} {\bibinfo {author} {\bibfnamefont {J.}~\bibnamefont
  {Tom}}, \bibinfo {author} {\bibfnamefont {M.}~\bibnamefont {Carbone}}, and\
  \bibinfo {author} {\bibfnamefont {A.~D.}\ \bibnamefont {Bragg}},\ }\bibfield
  {title} {\bibinfo {title} {Exploring the turbulent velocity gradients at
  different scales from the perspective of the strain-rate eigenframe},\
  }\href@noop {} {\bibfield  {journal} {\bibinfo  {journal} {J. Fluid Mech.}\
  }\textbf {\bibinfo {volume} {910}} (\bibinfo {year} {2021})}\BibitemShut
  {NoStop}%
\bibitem [{\citenamefont {Carbone}\ \emph {et~al.}(2020)\citenamefont
  {Carbone}, \citenamefont {Iovieno},\ and\ \citenamefont
  {Bragg}}]{carbone2020symmetry}%
  \BibitemOpen
  \bibfield  {author} {\bibinfo {author} {\bibfnamefont {M.}~\bibnamefont
  {Carbone}}, \bibinfo {author} {\bibfnamefont {M.}~\bibnamefont {Iovieno}},
  and\ \bibinfo {author} {\bibfnamefont {A.}~\bibnamefont {Bragg}},\ }\bibfield
   {title} {\bibinfo {title} {Symmetry transformation and dimensionality
  reduction of the anisotropic pressure hessian},\ }\href@noop {} {\bibfield
  {journal} {\bibinfo  {journal} {J. Fluid Mech.}\ }\textbf {\bibinfo {volume}
  {900}} (\bibinfo {year} {2020})}\BibitemShut {NoStop}%
\bibitem [{\citenamefont {Vieillefosse}(1982)}]{vieillefosse1982local}%
  \BibitemOpen
  \bibfield  {author} {\bibinfo {author} {\bibfnamefont {P.}~\bibnamefont
  {Vieillefosse}},\ }\bibfield  {title} {\bibinfo {title} {Local interaction
  between vorticity and shear in a perfect incompressible fluid},\ }\href@noop
  {} {\bibfield  {journal} {\bibinfo  {journal} {J. Phys.}\ }\textbf {\bibinfo
  {volume} {43}},\ \bibinfo {pages} {837} (\bibinfo {year} {1982})}\BibitemShut
  {NoStop}%
\bibitem [{\citenamefont {Vieillefosse}(1984)}]{vieillefosse1984internal}%
  \BibitemOpen
  \bibfield  {author} {\bibinfo {author} {\bibfnamefont {P.}~\bibnamefont
  {Vieillefosse}},\ }\bibfield  {title} {\bibinfo {title} {Internal motion of a
  small element of fluid in an inviscid flow},\ }\href@noop {} {\bibfield
  {journal} {\bibinfo  {journal} {Phys. A}\ }\textbf {\bibinfo {volume}
  {125}},\ \bibinfo {pages} {150} (\bibinfo {year} {1984})}\BibitemShut
  {NoStop}%
\bibitem [{\citenamefont {Cantwell}(1992)}]{cantwell1992exact}%
  \BibitemOpen
  \bibfield  {author} {\bibinfo {author} {\bibfnamefont {B.~J.}\ \bibnamefont
  {Cantwell}},\ }\bibfield  {title} {\bibinfo {title} {Exact solution of a
  restricted euler equation for the velocity gradient tensor},\ }\href@noop {}
  {\bibfield  {journal} {\bibinfo  {journal} {Phys. Fluids A: Fluid Dynam.}\
  }\textbf {\bibinfo {volume} {4}},\ \bibinfo {pages} {782} (\bibinfo {year}
  {1992})}\BibitemShut {NoStop}%
\bibitem [{\citenamefont {Martin}\ \emph
  {et~al.}(1998{\natexlab{b}})\citenamefont {Martin}, \citenamefont {Dopazo},\
  and\ \citenamefont {Vali{\~n}o}}]{martin1998diffusion}%
  \BibitemOpen
  \bibfield  {author} {\bibinfo {author} {\bibfnamefont {J.}~\bibnamefont
  {Martin}}, \bibinfo {author} {\bibfnamefont {C.}~\bibnamefont {Dopazo}}, and\
  \bibinfo {author} {\bibfnamefont {L.}~\bibnamefont {Vali{\~n}o}},\ }\bibfield
   {title} {\bibinfo {title} {Dynamics of velocity gradient invariants in
  turbulence: Restricted euler and linear diffusion models},\ }\href@noop {}
  {\bibfield  {journal} {\bibinfo  {journal} {Phys. Fluids}\ }\textbf {\bibinfo
  {volume} {10}},\ \bibinfo {pages} {2012} (\bibinfo {year}
  {1998}{\natexlab{b}})}\BibitemShut {NoStop}%
\bibitem [{\citenamefont {Girimaji}\ and\ \citenamefont
  {Pope}(1990)}]{girimaji1990diffusion}%
  \BibitemOpen
  \bibfield  {author} {\bibinfo {author} {\bibfnamefont {S.~S.}\ \bibnamefont
  {Girimaji}}and\ \bibinfo {author} {\bibfnamefont {S.~B.}\ \bibnamefont
  {Pope}},\ }\bibfield  {title} {\bibinfo {title} {A diffusion model for
  velocity gradients in turbulence},\ }\href@noop {} {\bibfield  {journal}
  {\bibinfo  {journal} {Phys. Fluids A: Fluid Dynam.}\ }\textbf {\bibinfo
  {volume} {2}},\ \bibinfo {pages} {242} (\bibinfo {year} {1990})}\BibitemShut
  {NoStop}%
\bibitem [{\citenamefont {Chertkov}\ \emph {et~al.}(1999)\citenamefont
  {Chertkov}, \citenamefont {Pumir},\ and\ \citenamefont
  {Shraiman}}]{chertkov1999lagrangian}%
  \BibitemOpen
  \bibfield  {author} {\bibinfo {author} {\bibfnamefont {M.}~\bibnamefont
  {Chertkov}}, \bibinfo {author} {\bibfnamefont {A.}~\bibnamefont {Pumir}},
  and\ \bibinfo {author} {\bibfnamefont {B.~I.}\ \bibnamefont {Shraiman}},\
  }\bibfield  {title} {\bibinfo {title} {Lagrangian tetrad dynamics and the
  phenomenology of turbulence},\ }\href@noop {} {\bibfield  {journal} {\bibinfo
   {journal} {Phys. Fluids}\ }\textbf {\bibinfo {volume} {11}},\ \bibinfo
  {pages} {2394} (\bibinfo {year} {1999})}\BibitemShut {NoStop}%
\bibitem [{\citenamefont {Jeong}\ and\ \citenamefont
  {Girimaji}(2003)}]{jeong2003velocity}%
  \BibitemOpen
  \bibfield  {author} {\bibinfo {author} {\bibfnamefont {E.}~\bibnamefont
  {Jeong}}and\ \bibinfo {author} {\bibfnamefont {S.~S.}\ \bibnamefont
  {Girimaji}},\ }\bibfield  {title} {\bibinfo {title} {Velocity-gradient
  dynamics in turbulence: effect of viscosity and forcing},\ }\href@noop {}
  {\bibfield  {journal} {\bibinfo  {journal} {Theor. Comp. Fluid Dyn.}\
  }\textbf {\bibinfo {volume} {16}},\ \bibinfo {pages} {421} (\bibinfo {year}
  {2003})}\BibitemShut {NoStop}%
\bibitem [{\citenamefont {Biferale}\ \emph {et~al.}(2007)\citenamefont
  {Biferale}, \citenamefont {Chevillard}, \citenamefont {Meneveau},\ and\
  \citenamefont {Toschi}}]{biferale2007multiscale}%
  \BibitemOpen
  \bibfield  {author} {\bibinfo {author} {\bibfnamefont {L.}~\bibnamefont
  {Biferale}}, \bibinfo {author} {\bibfnamefont {L.}~\bibnamefont
  {Chevillard}}, \bibinfo {author} {\bibfnamefont {C.}~\bibnamefont
  {Meneveau}}, and\ \bibinfo {author} {\bibfnamefont {F.}~\bibnamefont
  {Toschi}},\ }\bibfield  {title} {\bibinfo {title} {Multiscale model of
  gradient evolution in turbulent flows},\ }\href@noop {} {\bibfield  {journal}
  {\bibinfo  {journal} {Phys. Rev. Lett.}\ }\textbf {\bibinfo {volume} {98}},\
  \bibinfo {pages} {214501} (\bibinfo {year} {2007})}\BibitemShut {NoStop}%
\bibitem [{\citenamefont {Chevillard}\ and\ \citenamefont
  {Meneveau}(2006)}]{chevillard2006lagrangian}%
  \BibitemOpen
  \bibfield  {author} {\bibinfo {author} {\bibfnamefont {L.}~\bibnamefont
  {Chevillard}}and\ \bibinfo {author} {\bibfnamefont {C.}~\bibnamefont
  {Meneveau}},\ }\bibfield  {title} {\bibinfo {title} {Lagrangian dynamics and
  statistical geometric structure of turbulence},\ }\href@noop {} {\bibfield
  {journal} {\bibinfo  {journal} {Phys. Rev. Lett.}\ }\textbf {\bibinfo
  {volume} {97}},\ \bibinfo {pages} {174501} (\bibinfo {year}
  {2006})}\BibitemShut {NoStop}%
\bibitem [{\citenamefont {Chevillard}\ \emph {et~al.}(2008)\citenamefont
  {Chevillard}, \citenamefont {Meneveau}, \citenamefont {Biferale},\ and\
  \citenamefont {Toschi}}]{chevillard2008modeling}%
  \BibitemOpen
  \bibfield  {author} {\bibinfo {author} {\bibfnamefont {L.}~\bibnamefont
  {Chevillard}}, \bibinfo {author} {\bibfnamefont {C.}~\bibnamefont
  {Meneveau}}, \bibinfo {author} {\bibfnamefont {L.}~\bibnamefont {Biferale}},
  and\ \bibinfo {author} {\bibfnamefont {F.}~\bibnamefont {Toschi}},\
  }\bibfield  {title} {\bibinfo {title} {Modeling the pressure hessian and
  viscous laplacian in turbulence: comparisons with direct numerical simulation
  and implications on velocity gradient dynamics},\ }\href@noop {} {\bibfield
  {journal} {\bibinfo  {journal} {Phys. Fluids}\ }\textbf {\bibinfo {volume}
  {20}},\ \bibinfo {pages} {101504} (\bibinfo {year} {2008})}\BibitemShut
  {NoStop}%
\bibitem [{\citenamefont {Wilczek}\ and\ \citenamefont
  {Meneveau}(2014)}]{wilczek2014pressure}%
  \BibitemOpen
  \bibfield  {author} {\bibinfo {author} {\bibfnamefont {M.}~\bibnamefont
  {Wilczek}}and\ \bibinfo {author} {\bibfnamefont {C.}~\bibnamefont
  {Meneveau}},\ }\bibfield  {title} {\bibinfo {title} {Pressure hessian and
  viscous contributions to velocity gradient statistics based on gaussian
  random fields},\ }\href@noop {} {\bibfield  {journal} {\bibinfo  {journal}
  {J. Fluid Mech.}\ }\textbf {\bibinfo {volume} {756}},\ \bibinfo {pages} {191}
  (\bibinfo {year} {2014})}\BibitemShut {NoStop}%
\bibitem [{\citenamefont {Johnson}\ and\ \citenamefont
  {Meneveau}(2016)}]{johnson2016closure}%
  \BibitemOpen
  \bibfield  {author} {\bibinfo {author} {\bibfnamefont {P.~L.}\ \bibnamefont
  {Johnson}}and\ \bibinfo {author} {\bibfnamefont {C.}~\bibnamefont
  {Meneveau}},\ }\bibfield  {title} {\bibinfo {title} {A closure for lagrangian
  velocity gradient evolution in turbulence using recent-deformation mapping of
  initially gaussian fields},\ }\href@noop {} {\bibfield  {journal} {\bibinfo
  {journal} {J. Fluid Mech.}\ }\textbf {\bibinfo {volume} {804}},\ \bibinfo
  {pages} {387} (\bibinfo {year} {2016})}\BibitemShut {NoStop}%
\bibitem [{\citenamefont {Lawson}\ and\ \citenamefont
  {Dawson}(2015)}]{lawson2015velocity}%
  \BibitemOpen
  \bibfield  {author} {\bibinfo {author} {\bibfnamefont {J.~M.}\ \bibnamefont
  {Lawson}}and\ \bibinfo {author} {\bibfnamefont {J.~R.}\ \bibnamefont
  {Dawson}},\ }\bibfield  {title} {\bibinfo {title} {On velocity gradient
  dynamics and turbulent structure},\ }\href@noop {} {\bibfield  {journal}
  {\bibinfo  {journal} {J. Fluid Mech.}\ }\textbf {\bibinfo {volume} {780}},\
  \bibinfo {pages} {60} (\bibinfo {year} {2015})}\BibitemShut {NoStop}%
\bibitem [{\citenamefont {Leppin}\ and\ \citenamefont
  {Wilczek}(2020)}]{leppin2020capturing}%
  \BibitemOpen
  \bibfield  {author} {\bibinfo {author} {\bibfnamefont {L.~A.}\ \bibnamefont
  {Leppin}}and\ \bibinfo {author} {\bibfnamefont {M.}~\bibnamefont {Wilczek}},\
  }\bibfield  {title} {\bibinfo {title} {Capturing velocity gradients and
  particle rotation rates in turbulence},\ }\href@noop {} {\bibfield  {journal}
  {\bibinfo  {journal} {Phys. Rev Lett.}\ }\textbf {\bibinfo {volume} {125}},\
  \bibinfo {pages} {224501} (\bibinfo {year} {2020})}\BibitemShut {NoStop}%
\bibitem [{\citenamefont {Betchov}(1956)}]{betchov1956inequality}%
  \BibitemOpen
  \bibfield  {author} {\bibinfo {author} {\bibfnamefont {R.}~\bibnamefont
  {Betchov}},\ }\bibfield  {title} {\bibinfo {title} {An inequality concerning
  the production of vorticity in isotropic turbulence},\ }\href@noop {}
  {\bibfield  {journal} {\bibinfo  {journal} {Journal of Fluid Mechanics}\
  }\textbf {\bibinfo {volume} {1}},\ \bibinfo {pages} {497} (\bibinfo {year}
  {1956})}\BibitemShut {NoStop}%
\bibitem [{\citenamefont {Pereira}\ \emph {et~al.}(2018)\citenamefont
  {Pereira}, \citenamefont {Moriconi},\ and\ \citenamefont
  {Chevillard}}]{pereira2018multifractal}%
  \BibitemOpen
  \bibfield  {author} {\bibinfo {author} {\bibfnamefont {R.~M.}\ \bibnamefont
  {Pereira}}, \bibinfo {author} {\bibfnamefont {L.}~\bibnamefont {Moriconi}},
  and\ \bibinfo {author} {\bibfnamefont {L.}~\bibnamefont {Chevillard}},\
  }\bibfield  {title} {\bibinfo {title} {A multifractal model for the velocity
  gradient dynamics in turbulent flows},\ }\href@noop {} {\bibfield  {journal}
  {\bibinfo  {journal} {J. Fluid Mech.}\ }\textbf {\bibinfo {volume} {839}},\
  \bibinfo {pages} {430} (\bibinfo {year} {2018})}\BibitemShut {NoStop}%
\bibitem [{\citenamefont {Maulik}\ \emph {et~al.}(2019)\citenamefont {Maulik},
  \citenamefont {San}, \citenamefont {Rasheed},\ and\ \citenamefont
  {Vedula}}]{maulik2019subgrid}%
  \BibitemOpen
  \bibfield  {author} {\bibinfo {author} {\bibfnamefont {R.}~\bibnamefont
  {Maulik}}, \bibinfo {author} {\bibfnamefont {O.}~\bibnamefont {San}},
  \bibinfo {author} {\bibfnamefont {A.}~\bibnamefont {Rasheed}}, and\ \bibinfo
  {author} {\bibfnamefont {P.}~\bibnamefont {Vedula}},\ }\bibfield  {title}
  {\bibinfo {title} {Subgrid modelling for two-dimensional turbulence using
  neural networks},\ }\href@noop {} {\bibfield  {journal} {\bibinfo  {journal}
  {Journal of Fluid Mechanics}\ }\textbf {\bibinfo {volume} {858}},\ \bibinfo
  {pages} {122} (\bibinfo {year} {2019})}\BibitemShut {NoStop}%
\bibitem [{\citenamefont {Duraisamy}\ \emph {et~al.}(2019)\citenamefont
  {Duraisamy}, \citenamefont {Iaccarino},\ and\ \citenamefont
  {Xiao}}]{duraisamy2019turbulence}%
  \BibitemOpen
  \bibfield  {author} {\bibinfo {author} {\bibfnamefont {K.}~\bibnamefont
  {Duraisamy}}, \bibinfo {author} {\bibfnamefont {G.}~\bibnamefont
  {Iaccarino}}, and\ \bibinfo {author} {\bibfnamefont {H.}~\bibnamefont
  {Xiao}},\ }\bibfield  {title} {\bibinfo {title} {Turbulence modeling in the
  age of data},\ }\href@noop {} {\bibfield  {journal} {\bibinfo  {journal}
  {Annual Review of Fluid Mechanics}\ }\textbf {\bibinfo {volume} {51}},\
  \bibinfo {pages} {357} (\bibinfo {year} {2019})}\BibitemShut {NoStop}%
\bibitem [{\citenamefont {{King}}\ \emph {et~al.}(2018)\citenamefont {{King}},
  \citenamefont {{Hennigh}}, \citenamefont {{Mohan}},\ and\ \citenamefont
  {{Chertkov}}}]{2018PIML-LANL}%
  \BibitemOpen
  \bibfield  {author} {\bibinfo {author} {\bibfnamefont {R.}~\bibnamefont
  {{King}}}, \bibinfo {author} {\bibfnamefont {O.}~\bibnamefont {{Hennigh}}},
  \bibinfo {author} {\bibfnamefont {A.}~\bibnamefont {{Mohan}}}, and\ \bibinfo
  {author} {\bibfnamefont {M.}~\bibnamefont {{Chertkov}}},\ }\bibfield  {title}
  {\bibinfo {title} {{From Deep to Physics-Informed Learning of Turbulence:
  Diagnostics}},\ }\href@noop {} {\bibfield  {journal} {\bibinfo  {journal}
  {APS/DFD and arXiv:1810.07785}\ } (\bibinfo {year} {2018})}\BibitemShut
  {NoStop}%
\bibitem [{\citenamefont {Wang}\ \emph {et~al.}(2017)\citenamefont {Wang},
  \citenamefont {Wu},\ and\ \citenamefont {Xiao}}]{wang2017physics}%
  \BibitemOpen
  \bibfield  {author} {\bibinfo {author} {\bibfnamefont {J.-X.}\ \bibnamefont
  {Wang}}, \bibinfo {author} {\bibfnamefont {J.-L.}\ \bibnamefont {Wu}}, and\
  \bibinfo {author} {\bibfnamefont {H.}~\bibnamefont {Xiao}},\ }\bibfield
  {title} {\bibinfo {title} {Physics-informed machine learning approach for
  reconstructing reynolds stress modeling discrepancies based on dns data},\
  }\href@noop {} {\bibfield  {journal} {\bibinfo  {journal} {Physical Review
  Fluids}\ }\textbf {\bibinfo {volume} {2}},\ \bibinfo {pages} {034603}
  (\bibinfo {year} {2017})}\BibitemShut {NoStop}%
\bibitem [{\citenamefont {Mohan}\ \emph {et~al.}(2019)\citenamefont {Mohan},
  \citenamefont {Daniel}, \citenamefont {Chertkov},\ and\ \citenamefont
  {Livescu}}]{mohan2019compressed}%
  \BibitemOpen
  \bibfield  {author} {\bibinfo {author} {\bibfnamefont {A.}~\bibnamefont
  {Mohan}}, \bibinfo {author} {\bibfnamefont {D.}~\bibnamefont {Daniel}},
  \bibinfo {author} {\bibfnamefont {M.}~\bibnamefont {Chertkov}}, and\ \bibinfo
  {author} {\bibfnamefont {D.}~\bibnamefont {Livescu}},\ }\bibfield  {title}
  {\bibinfo {title} {Compressed convolutional lstm: An efficient deep learning
  framework to model high fidelity 3d turbulence},\ }\href@noop {} {\bibfield
  {journal} {\bibinfo  {journal} {arXiv preprint arXiv:1903.00033}\ } (\bibinfo
  {year} {2019})}\BibitemShut {NoStop}%
\bibitem [{\citenamefont {Fukami}\ \emph {et~al.}(2019)\citenamefont {Fukami},
  \citenamefont {Fukagata},\ and\ \citenamefont {Taira}}]{fukami2019super}%
  \BibitemOpen
  \bibfield  {author} {\bibinfo {author} {\bibfnamefont {K.}~\bibnamefont
  {Fukami}}, \bibinfo {author} {\bibfnamefont {K.}~\bibnamefont {Fukagata}},
  and\ \bibinfo {author} {\bibfnamefont {K.}~\bibnamefont {Taira}},\ }\bibfield
   {title} {\bibinfo {title} {Super-resolution reconstruction of turbulent
  flows with machine learning},\ }\href@noop {} {\bibfield  {journal} {\bibinfo
   {journal} {J. Fluid. Mech.}\ }\textbf {\bibinfo {volume} {870}},\ \bibinfo
  {pages} {106} (\bibinfo {year} {2019})}\BibitemShut {NoStop}%
\bibitem [{\citenamefont {Portwood}\ \emph {et~al.}(2019)\citenamefont
  {Portwood}, \citenamefont {Mitra}, \citenamefont {Ribeiro}, \citenamefont
  {Nguyen}, \citenamefont {Nadiga}, \citenamefont {Saenz}, \citenamefont
  {Chertkov}, \citenamefont {Garg}, \citenamefont {Anandkumar}, \citenamefont
  {Dengel} \emph {et~al.}}]{portwood2019turbulence}%
  \BibitemOpen
  \bibfield  {author} {\bibinfo {author} {\bibfnamefont {G.~D.}\ \bibnamefont
  {Portwood}}, \bibinfo {author} {\bibfnamefont {P.~P.}\ \bibnamefont {Mitra}},
  \bibinfo {author} {\bibfnamefont {M.~D.}\ \bibnamefont {Ribeiro}}, \bibinfo
  {author} {\bibfnamefont {T.~M.}\ \bibnamefont {Nguyen}}, \bibinfo {author}
  {\bibfnamefont {B.~T.}\ \bibnamefont {Nadiga}}, \bibinfo {author}
  {\bibfnamefont {J.~A.}\ \bibnamefont {Saenz}}, \bibinfo {author}
  {\bibfnamefont {M.}~\bibnamefont {Chertkov}}, \bibinfo {author}
  {\bibfnamefont {A.}~\bibnamefont {Garg}}, \bibinfo {author} {\bibfnamefont
  {A.}~\bibnamefont {Anandkumar}}, \bibinfo {author} {\bibfnamefont
  {A.}~\bibnamefont {Dengel}},  \emph {et~al.},\ }\bibfield  {title} {\bibinfo
  {title} {Turbulence forecasting via neural ode},\ }\href@noop {} {\bibfield
  {journal} {\bibinfo  {journal} {arXiv preprint arXiv:1911.05180}\ } (\bibinfo
  {year} {2019})}\BibitemShut {NoStop}%
\bibitem [{\citenamefont {Ohkitani}\ and\ \citenamefont
  {Kishiba}(1995)}]{ohkitani1995nonlocal}%
  \BibitemOpen
  \bibfield  {author} {\bibinfo {author} {\bibfnamefont {K.}~\bibnamefont
  {Ohkitani}}and\ \bibinfo {author} {\bibfnamefont {S.}~\bibnamefont
  {Kishiba}},\ }\bibfield  {title} {\bibinfo {title} {Nonlocal nature of vortex
  stretching in an inviscid fluid},\ }\href@noop {} {\bibfield  {journal}
  {\bibinfo  {journal} {Phys. Fluids}\ }\textbf {\bibinfo {volume} {7}},\
  \bibinfo {pages} {411} (\bibinfo {year} {1995})}\BibitemShut {NoStop}%
\bibitem [{\citenamefont {Pope}(1975)}]{pope1975more}%
  \BibitemOpen
  \bibfield  {author} {\bibinfo {author} {\bibfnamefont {S.~B.}\ \bibnamefont
  {Pope}},\ }\bibfield  {title} {\bibinfo {title} {A more general
  effective-viscosity hypothesis},\ }\href@noop {} {\bibfield  {journal}
  {\bibinfo  {journal} {J. Fluid Mech.}\ }\textbf {\bibinfo {volume} {72}},\
  \bibinfo {pages} {331} (\bibinfo {year} {1975})}\BibitemShut {NoStop}%
\bibitem [{\citenamefont {Zheng}(1993)}]{zheng1993representations}%
  \BibitemOpen
  \bibfield  {author} {\bibinfo {author} {\bibfnamefont {Q.-S.}\ \bibnamefont
  {Zheng}},\ }\bibfield  {title} {\bibinfo {title} {On the representations for
  isotropic vector-valued, symmetric tensor-valued and skew-symmetric
  tensor-valued functions},\ }\href@noop {} {\bibfield  {journal} {\bibinfo
  {journal} {International journal of engineering science}\ }\textbf {\bibinfo
  {volume} {31}},\ \bibinfo {pages} {1013} (\bibinfo {year}
  {1993})}\BibitemShut {NoStop}%
\bibitem [{\citenamefont {Hornik}(1991)}]{hornik1991approximation}%
  \BibitemOpen
  \bibfield  {author} {\bibinfo {author} {\bibfnamefont {K.}~\bibnamefont
  {Hornik}},\ }\bibfield  {title} {\bibinfo {title} {Approximation capabilities
  of multilayer feedforward networks},\ }\href@noop {} {\bibfield  {journal}
  {\bibinfo  {journal} {Neural networks}\ }\textbf {\bibinfo {volume} {4}},\
  \bibinfo {pages} {251} (\bibinfo {year} {1991})}\BibitemShut {NoStop}%
\bibitem [{\citenamefont {Kingma}\ and\ \citenamefont
  {Ba}(2014)}]{kingma2014adam}%
  \BibitemOpen
  \bibfield  {author} {\bibinfo {author} {\bibfnamefont {D.~P.}\ \bibnamefont
  {Kingma}}and\ \bibinfo {author} {\bibfnamefont {J.}~\bibnamefont {Ba}},\
  }\bibfield  {title} {\bibinfo {title} {Adam: A method for stochastic
  optimization},\ }\href@noop {} {\bibfield  {journal} {\bibinfo  {journal}
  {arXiv preprint arXiv:1412.6980}\ } (\bibinfo {year} {2014})}\BibitemShut
  {NoStop}%
\bibitem [{\citenamefont {Petersen}\ and\ \citenamefont
  {Livescu}(2010)}]{petersen2010forcing}%
  \BibitemOpen
  \bibfield  {author} {\bibinfo {author} {\bibfnamefont {M.}~\bibnamefont
  {Petersen}}and\ \bibinfo {author} {\bibfnamefont {D.}~\bibnamefont
  {Livescu}},\ }\bibfield  {title} {\bibinfo {title} {Forcing for statistically
  stationary compressible isotropic turbulence},\ }\href@noop {} {\bibfield
  {journal} {\bibinfo  {journal} {Physics of Fluids}\ }\textbf {\bibinfo
  {volume} {22}},\ \bibinfo {pages} {116101} (\bibinfo {year}
  {2010})}\BibitemShut {NoStop}%
\bibitem [{\citenamefont {Daniel}\ \emph {et~al.}(2018)\citenamefont {Daniel},
  \citenamefont {Livescu},\ and\ \citenamefont {Ryu}}]{daniel2018reaction}%
  \BibitemOpen
  \bibfield  {author} {\bibinfo {author} {\bibfnamefont {D.}~\bibnamefont
  {Daniel}}, \bibinfo {author} {\bibfnamefont {D.}~\bibnamefont {Livescu}},
  and\ \bibinfo {author} {\bibfnamefont {J.}~\bibnamefont {Ryu}},\ }\bibfield
  {title} {\bibinfo {title} {Reaction analogy based forcing for incompressible
  scalar turbulence},\ }\href@noop {} {\bibfield  {journal} {\bibinfo
  {journal} {Physical Review Fluids}\ }\textbf {\bibinfo {volume} {3}},\
  \bibinfo {pages} {094602} (\bibinfo {year} {2018})}\BibitemShut {NoStop}%
\bibitem [{\citenamefont {Yeung}\ and\ \citenamefont
  {Pope}(1988)}]{yeung1988algorithm}%
  \BibitemOpen
  \bibfield  {author} {\bibinfo {author} {\bibfnamefont {P.~K.}\ \bibnamefont
  {Yeung}}and\ \bibinfo {author} {\bibfnamefont {S.~B.}\ \bibnamefont {Pope}},\
  }\bibfield  {title} {\bibinfo {title} {An algorithm for tracking fluid
  particles in numerical simulations of homogeneous turbulence},\ }\href@noop
  {} {\bibfield  {journal} {\bibinfo  {journal} {J. Comput. Phys.}\ }\textbf
  {\bibinfo {volume} {79}},\ \bibinfo {pages} {373} (\bibinfo {year}
  {1988})}\BibitemShut {NoStop}%
\bibitem [{\citenamefont {Abadi}\ \emph {et~al.}(2015)\citenamefont {Abadi},
  \citenamefont {Agarwal}, \citenamefont {Barham}, \citenamefont {Brevdo},
  \citenamefont {Chen}, \citenamefont {Citro}, \citenamefont {Corrado},
  \citenamefont {Davis}, \citenamefont {Dean}, \citenamefont {Devin},
  \citenamefont {Ghemawat}, \citenamefont {Goodfellow}, \citenamefont {Harp},
  \citenamefont {Irving}, \citenamefont {Isard}, \citenamefont {Jia},
  \citenamefont {Jozefowicz}, \citenamefont {Kaiser}, \citenamefont {Kudlur},
  \citenamefont {Levenberg}, \citenamefont {Man\'{e}}, \citenamefont {Monga},
  \citenamefont {Moore}, \citenamefont {Murray}, \citenamefont {Olah},
  \citenamefont {Schuster}, \citenamefont {Shlens}, \citenamefont {Steiner},
  \citenamefont {Sutskever}, \citenamefont {Talwar}, \citenamefont {Tucker},
  \citenamefont {Vanhoucke}, \citenamefont {Vasudevan}, \citenamefont
  {Vi\'{e}gas}, \citenamefont {Vinyals}, \citenamefont {Warden}, \citenamefont
  {Wattenberg}, \citenamefont {Wicke}, \citenamefont {Yu},\ and\ \citenamefont
  {Zheng}}]{tensorflow2015-whitepaper}%
  \BibitemOpen
  \bibfield  {author} {\bibinfo {author} {\bibfnamefont {M.}~\bibnamefont
  {Abadi}}, \bibinfo {author} {\bibfnamefont {A.}~\bibnamefont {Agarwal}},
  \bibinfo {author} {\bibfnamefont {P.}~\bibnamefont {Barham}}, \bibinfo
  {author} {\bibfnamefont {E.}~\bibnamefont {Brevdo}}, \bibinfo {author}
  {\bibfnamefont {Z.}~\bibnamefont {Chen}}, \bibinfo {author} {\bibfnamefont
  {C.}~\bibnamefont {Citro}}, \bibinfo {author} {\bibfnamefont {G.~S.}\
  \bibnamefont {Corrado}}, \bibinfo {author} {\bibfnamefont {A.}~\bibnamefont
  {Davis}}, \bibinfo {author} {\bibfnamefont {J.}~\bibnamefont {Dean}},
  \bibinfo {author} {\bibfnamefont {M.}~\bibnamefont {Devin}}, \bibinfo
  {author} {\bibfnamefont {S.}~\bibnamefont {Ghemawat}}, \bibinfo {author}
  {\bibfnamefont {I.}~\bibnamefont {Goodfellow}}, \bibinfo {author}
  {\bibfnamefont {A.}~\bibnamefont {Harp}}, \bibinfo {author} {\bibfnamefont
  {G.}~\bibnamefont {Irving}}, \bibinfo {author} {\bibfnamefont
  {M.}~\bibnamefont {Isard}}, \bibinfo {author} {\bibfnamefont
  {Y.}~\bibnamefont {Jia}}, \bibinfo {author} {\bibfnamefont {R.}~\bibnamefont
  {Jozefowicz}}, \bibinfo {author} {\bibfnamefont {L.}~\bibnamefont {Kaiser}},
  \bibinfo {author} {\bibfnamefont {M.}~\bibnamefont {Kudlur}}, \bibinfo
  {author} {\bibfnamefont {J.}~\bibnamefont {Levenberg}}, \bibinfo {author}
  {\bibfnamefont {D.}~\bibnamefont {Man\'{e}}}, \bibinfo {author}
  {\bibfnamefont {R.}~\bibnamefont {Monga}}, \bibinfo {author} {\bibfnamefont
  {S.}~\bibnamefont {Moore}}, \bibinfo {author} {\bibfnamefont
  {D.}~\bibnamefont {Murray}}, \bibinfo {author} {\bibfnamefont
  {C.}~\bibnamefont {Olah}}, \bibinfo {author} {\bibfnamefont {M.}~\bibnamefont
  {Schuster}}, \bibinfo {author} {\bibfnamefont {J.}~\bibnamefont {Shlens}},
  \bibinfo {author} {\bibfnamefont {B.}~\bibnamefont {Steiner}}, \bibinfo
  {author} {\bibfnamefont {I.}~\bibnamefont {Sutskever}}, \bibinfo {author}
  {\bibfnamefont {K.}~\bibnamefont {Talwar}}, \bibinfo {author} {\bibfnamefont
  {P.}~\bibnamefont {Tucker}}, \bibinfo {author} {\bibfnamefont
  {V.}~\bibnamefont {Vanhoucke}}, \bibinfo {author} {\bibfnamefont
  {V.}~\bibnamefont {Vasudevan}}, \bibinfo {author} {\bibfnamefont
  {F.}~\bibnamefont {Vi\'{e}gas}}, \bibinfo {author} {\bibfnamefont
  {O.}~\bibnamefont {Vinyals}}, \bibinfo {author} {\bibfnamefont
  {P.}~\bibnamefont {Warden}}, \bibinfo {author} {\bibfnamefont
  {M.}~\bibnamefont {Wattenberg}}, \bibinfo {author} {\bibfnamefont
  {M.}~\bibnamefont {Wicke}}, \bibinfo {author} {\bibfnamefont
  {Y.}~\bibnamefont {Yu}}, and\ \bibinfo {author} {\bibfnamefont
  {X.}~\bibnamefont {Zheng}},\ }\href {https://www.tensorflow.org/} {\bibinfo
  {title} {{TensorFlow}: Large-scale machine learning on heterogeneous
  systems}} (\bibinfo {year} {2015}),\ \bibinfo {note} {software available from
  tensorflow.org}\BibitemShut {NoStop}%
\bibitem [{\citenamefont {Chollet}\ \emph {et~al.}(2015)\citenamefont {Chollet}
  \emph {et~al.}}]{chollet2015keras}%
  \BibitemOpen
  \bibfield  {author} {\bibinfo {author} {\bibfnamefont {F.}~\bibnamefont
  {Chollet}} \emph {et~al.},\ }\href@noop {} {\bibinfo {title} {Keras}},\
  \bibinfo {howpublished} {\url{https://keras.io}} (\bibinfo {year}
  {2015})\BibitemShut {NoStop}%
\bibitem [{\citenamefont {Hyett}\ \emph {et~al.}(2020)\citenamefont {Hyett},
  \citenamefont {Chertkov}, \citenamefont {Tian},\ and\ \citenamefont
  {Livescu}}]{criston2020}%
  \BibitemOpen
  \bibfield  {author} {\bibinfo {author} {\bibfnamefont {C.}~\bibnamefont
  {Hyett}}, \bibinfo {author} {\bibfnamefont {M.}~\bibnamefont {Chertkov}},
  \bibinfo {author} {\bibfnamefont {Y.}~\bibnamefont {Tian}}, and\ \bibinfo
  {author} {\bibfnamefont {D.}~\bibnamefont {Livescu}},\ }\bibfield  {title}
  {\bibinfo {title} {Machine learning statistical lagrangian geometry of
  turbulence},\ }\href@noop {} {\bibfield  {journal} {\bibinfo  {journal}
  {APS}\ } (\bibinfo {year} {2020})}\BibitemShut {NoStop}%
\end{thebibliography}%

\end{document}